%% file: ms.tex
\documentclass[sigconf]{acmart}

\setcopyright{none}
\renewcommand\footnotetextcopyrightpermission[1]{}
\usepackage{multirow,makecell}

\usepackage{amssymb}
\usepackage{xcolor}

\newif\ifshowcomments
\showcommentsfalse

\ifshowcomments
\newcommand{\mynote}[2]{\fbox{\bfseries\sffamily\scriptsize{#1}}
 {\small$\blacktriangleright$\textsf{\emph{#2}}$\blacktriangleleft$}}
\else
\newcommand{\mynote}[2]{}
\fi

\newcommand{\ClassSet}{\mathcal{C}}
\newcommand{\FlowTimeout}{T_F}
\newcommand{\ActivityTimeout}{T_A}

\begin{document}

%
\title{Peel the onion: Recognition of Android apps behind the Tor Network}

%

\author{Emanuele Petagna}
\email{petagna.795137@studenti.uniroma1.it}
\affiliation{%
  \institution{Department of Computer, Control, and Management Engineering, ``Antonio Ruberti'' (DIAG)}
  \streetaddress{Via Ariosto, 25}
  \city{Rome}
  \country{Italy}
}

\author{Giuseppe Laurenza}
\email{laurenza@diag.uniroma1.it}
\affiliation{%
  \institution{Department of Computer, Control, and Management Engineering, ``Antonio Ruberti'' (DIAG)}
  \streetaddress{Via Ariosto, 25}
  \city{Rome}
  \country{Italy}
}

\author{Claudio Ciccotelli}
\email{ciccotelli@diag.uniroma1.it}
\affiliation{%
  \institution{Department of Computer, Control, and Management Engineering, ``Antonio Ruberti'' (DIAG)}
  \streetaddress{Via Ariosto, 25}
  \city{Rome}
  \country{Italy}
}

\author{Leonardo Querzoni}
\email{querzoni@diag.uniroma1.it}
\affiliation{%
  \institution{Department of Computer, Control, and Management Engineering, ``Antonio Ruberti'' (DIAG)}
  \streetaddress{Via Ariosto, 25}
  \city{Rome}
  \country{Italy}
}

%
%
%
%
%

%
\renewcommand{\shortauthors}{Petagna, et al.}

%
\begin{abstract}
In this work we show that Tor is vulnerable to app deanonymization attacks on Android devices through network traffic analysis.
For this purpose, we describe a general methodology for performing an attack that allows to deanonymize the apps running on a target smartphone using Tor, which is the victim of the attack.
Then, we discuss a Proof-of-Concept, implementing the methodology, that shows how the attack can be performed in practice and allows to assess the deanonymization accuracy that it is possible to achieve.
While attacks against Tor anonymity have been already gained considerable attention in the context of website fingerprinting in desktop environments, to the best of our knowledge this is the first work that highlights Tor vulnerability to apps deanonymization attacks on Android devices.
In our experiments we achieved an accuracy of $97\%$.
\end{abstract}

%
%
%

%

%

%
\maketitle

\input{introduction}
\input{related_works}
\input{background}
\input{threat_model}
\input{methodology}
\input{proof_of_concept}

\input{experiments}
\input{conclusion}

%
\bibliographystyle{plain}
\bibliography{ms}

%
\input{appendix}

\end{document}

%% file: introduction.tex
\section{Introduction}\label{sec:introduction}

Tor is a very popular anonymization network, currently counting more than two million of daily users~\cite{tormetrics}.
While in desktop environments Tor is mainly associated with preserving anonymity during Web navigation, its protection capabilities are not limited to such application.
In general, Tor can be used to protect any TCP-based traffic.
Nowadays, smartphone apps are replacing web browsers for interacting with many online services, such as social networks, chat services and video/audio streaming.
The usage of anonymization mechanisms, such as Tor, on mobile devices is motivated by the increasing interest in profiling mobile users, e.g., for marketing, governments surveillance, detection and exploitation of vulnerabilities and other activities that may be harmful for users' privacy and security, or perceived as such by them.

The aim of this work is showing that Tor is vulnerable to app deanonymization attacks on Android devices through network traffic analysis.
For this purpose, we describe a general methodology for performing an attack that allows to deanonymize the apps running on a target smartphone using Tor, which is the victim of the attack.
Then, we discuss a Proof-of-Concept, implementing the methodology, that shows how the attack can be performed in practice and allows to assess the deanonymization accuracy that it is possible to achieve.

Summarizing, this work provides the following contributions:
\begin{itemize}
\item A methodology for deanonymizing apps on Android-based smartphones behind Tor;
\item A Proof-of-Concept that implements the deanonymization methodology, which can be used to verify Tor's vulnerability to app deanonymization and assess the level of accuracy that can be achieved;
\item A dataset\footnote{Both the software necessary to reproduce the Proof-of-Concept and the dataset can be downloaded from the following repository \url{https://www.cis.uniroma1.it/peel-the-onion}} of generated Android Tor traffic traces that can be used to check the validity of our Proof-of-Concept and compare alternative methodologies.
\end{itemize}  

The remainder of the paper is organized as follows. 
Section~\ref{sec:related_works} reports the related works. 
Section~\ref{sec:background} presents the fundamental concepts related to Tor and the machine learning algorithms employed in this work.
Section~\ref{sec:threatmodel} introduces the threat model that we consider.
Section~\ref{sec:methodology} discusses the methodology for deanonymizing Android apps behind the Tor network.
Section~\ref{sec:proof-of-concept} describes the Proof-of-Concept.
Section~\ref{sec:exp-eval} reports the experiment performed to evaluate the accuracy of the methodology and discusses the obtained results.
Finally, in section~\ref{sec:conclusion} we draw some conclusion and discuss possible future directions for this work. 

%% file: related_works.tex
\section{Related Works}\label{sec:related_works}

Many works have been published in the broad area of traffic analysis both in the context of \emph{desktop environments} and \emph{smartphone environments} (mostly assuming the Android operating system). 
While, there are some works in the context of desktop environments that has focused on deanonymizing Tor traffic, to the best of our knowledge, there is no work assuming both a smartphone environment and that traffic is anonymized through Tor. 
Therefore, there is no work we can directly compare to. 

In this section we report the most related works considering a desktop environment, with or without Tor anonymized traffic, and an Android environment without Tor.

\subsection{Desktop environment without Tor}

In the context of website fingerprinting, Hintz~\cite{resourcelen} proposes an attack against SafeWeb, an encrypting web proxy, that allows to determine the webpages visited by the users.
The attack exploit the fact that, even if traffic is encrypted, many web browser open a separate TCP connection for downloading each resource in a visited webpage. 
This allows an attacker to monitor the size of the resources associated to a web page by analyzing the amount of bytes transferred in the corresponding TCP connections. 
The set of resources' size are used to fingerprint webpages.
The author also proposes some protection mechanisms against this attack, such as adding noise to traffic and using single connections to download all the resources of a webpage.

Bissias \emph{et al.}~\cite{biintetiming} propose a statistical website fingerprinting attack.  
The attacker creates a profile of the target website by monitoring the distribution of packet sizes and inter-arrival times. 
These data are then compared to user traffic. 
The authors show that the attack achieve an accuracy of 23\% on the entire dataset and an accuracy of 40\% on the set of top 25 most identifiable websites. 

Liberatore \emph{et al.}~\cite{liberatorenbayes} describe a website fingerprinting attack against HTTPS connections. 
They use unique packet lengths to build profiles of HTTPS connections and compare them against a dataset of known profiles using a naive Bayes classifier.

\subsection{Desktop environment with Tor}

In the context of Website fingerprinting, Wang \emph{et al.}~\cite{waknn} propose an attack that uses a $k-$Nearest Neighbor Classifier to effectively fingerprint web pages behind Tor.
They employ several types of features, including general statistics about total traffic (total incoming/outgoing packets, total transmission time, etc.), unique packet lengths, packet orderings, bursts (e.g., length of the first 100 burts, direction of the first 10 packets) and inter-packet times.
They show that their attack has significant higher accuracy than previous  attacks in the same field. 

AlSabah \emph{et al.}~\cite{realtimetor} propose a machine learning based approach for Tor's traffic classification.
The aim of the work is to recognize different classes of workloads that, in combination with QoS policies, can significantly improve the experience of Tor clients.
Indeed, as stated in the paper, most of the Tor's traffic is interactive web browsing but there is a small part of this traffic that consists of bulk download. 
This is a problem for a network that tries to be low-latency and that relies on a relatively small number of nodes. 
However, since Tor's traffic is encrypted, it is not possible to rely on classical QoS to discriminate applications traffic.
The proposed technique achieve an accuracy higher than 95\%.

Johnson~\emph{et al.}~\cite{routedtor} analyze Tor security in real world. The authors try to model various typical Tor users and examine different types of adversary models over the Internet network.
They define a methodology and metrics to measure the security of users over Tor.

Juarez~\emph{et al.}~\cite{torwf} analyzes the known website fingerprinting attacks on Tor. 
Known attacks claim to be effective under precise assumptions about threat model and user settings, which often do not hold in practical scenarios. 
The authors conduct a critical evaluation of these attacks and show their weakness when performed in real scenarios.
  
Chakravarty~\emph{et al.}~\cite{traffictorflow} evaluates the feasibility and effectiveness of practical traffic analysis attacks on the Tor network using NetFlow data. 
It is not a passive attack, authors deliberately alter traffic characteristic at the server side and observe how this alteration affects client side through a statistical correlation. 
They achieve 100\% accuracy in laboratory tests, and 81.4\% accuracy in real world tests.

Ling~\emph{et al.}~\cite{torward} propose TorWard, a system that attempts to recognize malicious traffic over Tor. Indeed, Tor anonymity is often abused to send malicious traffic (e.g., P2P botnets, trojan, etc.) with the aim of bypassing traditional IDS.
In their experiments they found that a considerable portion of the Tor traffic is malicious (around 10\%) with 8.99\% of the alerts generated due to malware and 78.03\% of the alerts generated due to malicious P2P traffic.

Mittal~\emph{et al.}~\cite{throoughputfingerptinting} exploits throughput information to gain information about the user.
The attack can identify the Guard Node (entry point to 
Tor network) and identify if two concurrent TCP connections belong to the same user.

Habibi Lashkari~\emph{et al.}~\cite{habibi} focus on recognition of traffic types instead of websites.
They consider are 8 application traffic types: browsing, email, chat, audio streaming, video streaming, file transfer, VoIP and P2P.
They perform network traffic analysis by splitting the traffic traces in flows of a certain duration.
For each flow they compute several features based on inter-arrival times, active and idle periods, packet rates and byte rates.
They employ a supervised machine learning approach to classify the traffic type of each flow. 
In particular they explored $k-$Nearest Neighbor, Random Forest and C4.5 classifiers.

\subsection{Android environment without Tor}

A number of authors have proposed various approaches to identify smartphone 
apps through network traffic analysis. 
Many of these solutions focus on examining IP addresses and packet payloads. 
However, relying on IP addresses is less effective because a lot of applications exploit Content Delivery Network (CDN) for scalability.

AppScanner~\cite{Appscanner} targets mobile environments and uses traffic features to fingerprint mobile apps.
They rely on a supervised machine learning approach using only features that does not require to inspect packet payloads, thus working also on encrypted traffic.
They perform experiments with SVM and Random Forest classifiers achieving 99\% of accuracy in their dataset with 110 of the most popular apps in the Google Play Store.

Dai~\emph{et al.}~\cite{androidfingerprinting1} propose a technique for app fingerprinting based on building network traffic profiles of apps.
They run each app in an emulator, exercising different execution paths through a novel UI fuzzing technique, and collect the corresponding network traces.
They compute a fingerprint of the app by identifying invariants in the generated network traces.
Using the generated fingerprint they were able to detect the presence of apps in real-world network traffic logs from a cellular provider.

Conti~\emph{et al.}~\cite{actionfingerprinting} describe a machine learning based network traffic analysis approach to identify user actions on specific apps (facebook, gmail and twitter). 
They achieve more than 95\% of accuracy and precision for most of the considered actions.

St{\"o}ber~\emph{et al.}~\cite{android3gsfingerprinting} focuses on identifying smartphones from 3G/UMTS data capture. 
Even if 3G/UMTS data is encrypted an attacker could reliably identify a smartphone using only the information extracted from periodic traffic patterns leak side-channel information like timing and data volume. 
They show as by computing fingerprints on 6 hours of background traffic sniffing, allows to reliably identify a smartphone, with a success probability of 90\%, with only about 15 minutes of traffic monitoring.

Saltaformaggio~\emph{et al.}~\cite{actionfingeprinintingIP} develop a tool called NetScope which is able to detect user activities on both Android and iOS smartphones. 
They compute feature by only inspecting the IP headers, and use a SVM multi-class classifier to detect activities.
NetScope achieves a precision of 78.04\% and a recall of 76.04\% on average
on a set of 35 widely used apps. 

%% file: background.tex
\section{Background on Tor}\label{sec:background}

In this section we briefly summarize the basic concepts about the Tor network.
Tor \cite{tordesign2014} is a distributed overlay network that anonymizes TCP-based applications (web browsers, secure shell, mail clients) while trying to keep the latency low. 
The network consists of a set of interconnected entities called \emph{Onion Routers} (ORs).
Tor clients, also known as \emph{Onion Proxies} (OPs), periodically connect to directory servers to download the list of available ORs.
OPs use this information to establish \emph{circuits} in the Tor network, to connect to a destination node (which is outside the Tor network).
A circuit is a path of ORs in which each OR knows only its predecessor and its successor OR.
A Tor circuit has three types of nodes:
\begin{itemize}
  \item \emph{Entry} or \emph{Guard Node}: this represents the entry point to Tor network for the Tor client. 
  
  \item \emph{Relay Nodes}: these are the intermediate ORs of the circuit.
  
  \item \emph{Exit Node}: this is the last node in the Tor circuit. That is, the one that connects to the destination.
\end{itemize}

Each Tor circuit must have one entry node, at least one relay node (but there may be multiple) and one exit node.
The entry node is the only node in the circuit that knows the Tor client, while the exit node is the only one that knows the destination.

Messages exchanged between the Tor client and the destination are split into \emph{cells} when they traverse the Tor network.
Cells are the basic unit of communication among Tor nodes.
Tor cells used to have a $512$ bytes fixed size in earlier Tor versions. 
Though this choice provided some resistance against traffic analysis, it was inefficient and made Tor traffic easier to discover due to packet-size distribution~\cite{tordesign2014}.
Therefore, variable length cells have been introduced in newer Tor versions.

\begin{figure*}[t]
  \centering
    \includegraphics[width=0.8\linewidth]{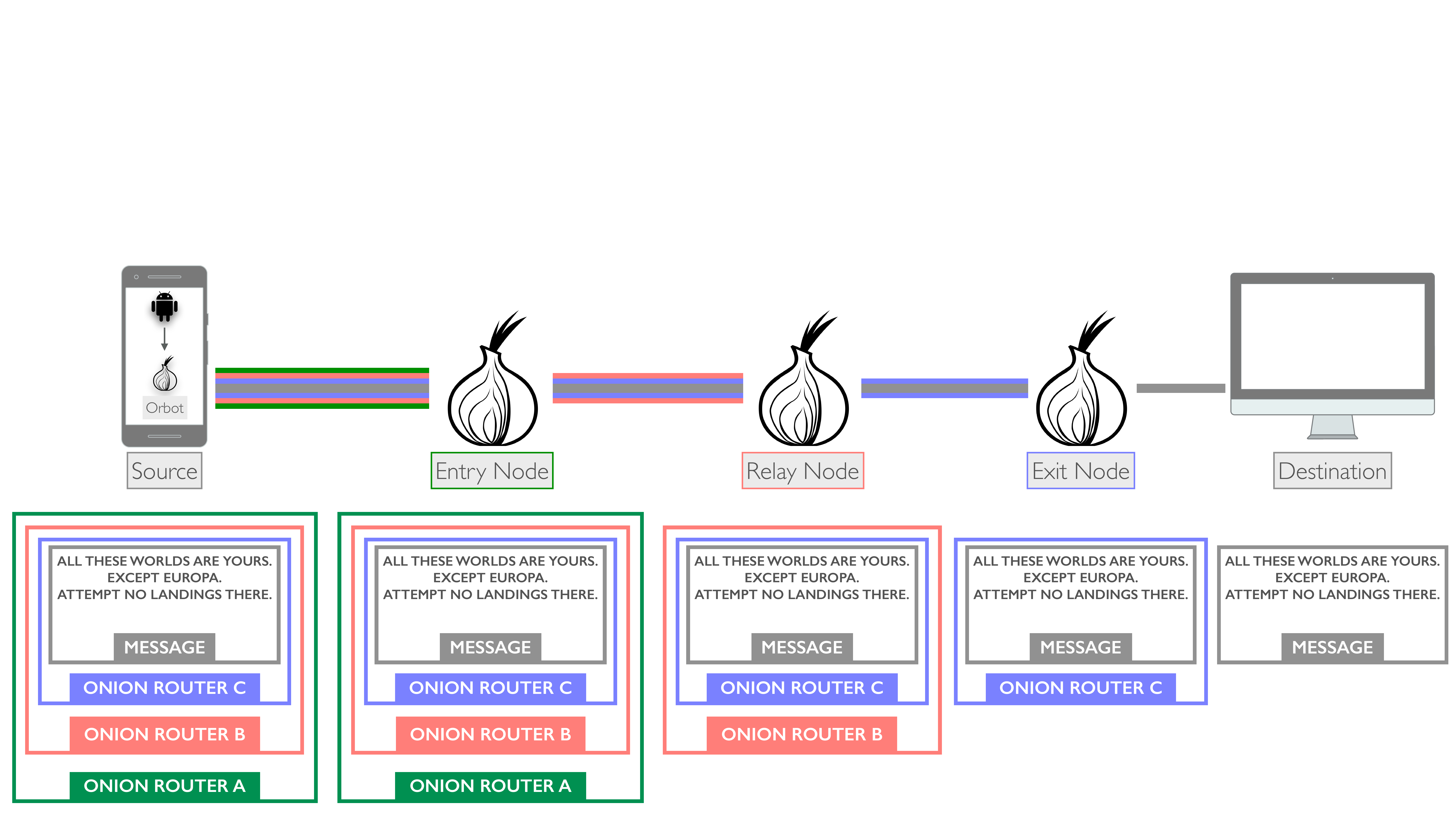}
  \caption{Cell encryption along the circuit from source to destination.}
  \Description{Cell encryption along the circuit from source to destination.}
  \label{fig:torsendpacket}   
\end{figure*}

When establishing a circuit, the Tor client shares a symmetric key with each node of the circuit.
As shown in Figure~\ref{fig:torsendpacket}, when the Tor client sends a packet to the destination it encrypts the corresponding cells' payloads with all the shared keys, in reverse order from the exit node to the entry node. 
Each node along the path unwraps its layer using its key.
Only the exit node can reconstruct the the message to be sent to the destination in clear.
On the opposite direction, when the destination sends a packet to the Tor client, each node in the circuit encrypts the corresponding cells, so that only the Tor client can reconstruct the original message by decrypting the cells using all the shared keys in the order from the entry to the exit node. 

\subsection{Padding}\label{sec:padding}
Internet service providers and surveillance infrastructures are known to store metadata about connections. 
Routers supports metadata retention through standard protocols (e.g., Netflow, jFlow, Netstream, IPFIX, etc.) typically in the form of per-flow records, containing at least the so called 5-tuple (consisting of the  source IP address, source port number, destination IP address, destination port number and protocol fields) and typically other information, such as the amount of bytes sent/received.
Collecting and analyzing such data is useful for characterizing traffic, but may also represent a threat to anonymity.

Per-flow records are emitted by routers on a periodic basis depending on two configurable timeouts: \emph{active flow timeout} and the \emph{inactive flow timeout}.
The expiration of the active flow timeout causes routers to emit a new record for each active connection. 
The inactive flow timeout causes the emission of a new record when a connection is inactive for a certain amount of time.
The value of such timeouts is configurable and the range depends on routers vendors, but active flow timeout is typically in the order of minutes, while the inactive flow timeout in the order of tens of seconds.
Therefore, the aggregation level of records data (on a temporal basis) is at least the active flow timeout, but may be finer when there are inactive periods longer than the inactive flow timeout. 

Thus, to reduce the granularity level of records' data (with the aim of hindering deanonymization techniques based on traffic analysis), long inactive periods should be avoided.
For this reason, the Tor protocol introduced \emph{connection padding}.
With connection padding, special purpose cells (PADDING cells) are sent if the connection is inactive for a given amount of time, so as to reduce the duration of inactive periods.

\subsubsection{Connection Padding}\label{sec:fullpadding}
Connection padding cells are exchanged only between the Tor client and entry node. No other nodes in the circuit are involved.
To determine when to send a connection padding cell, both the Tor client and the entry node maintain a timer. 
These timers are set up with a timeout value between $1.5$ and $9.5$ seconds. 
The exact value depends on a function that samples a distribution described in \cite{torpadding}.
After the establishment of the Tor circuit the timers start on both sides, if any of the two timers expires, a padding cell is sent to the other endpoint.
Exchanging any cell different from a padding cell resets the timers.

\subsubsection{Reduced Connection Padding}\label{sec:reducedpadding}
Connection padding introduces an overhead in terms of exchanged data.
Especially in mobile environments, where clients usually have limited data 
plan and reducing traffic helps saving battery life, this overhead may become excessive.
Therefore, \emph{reduced connection padding} has been introduced to lower the overhead due to connection padding. 
With reduced connection padding the timeout is sampled from a different range, between $9$ seconds to $14$ seconds.
This results in a less frequent exchange of padding cells with respect to connection padding.

%% file: threat_model.tex
\section{Threat Model}
\label{sec:threatmodel}

\begin{figure}[t]
  \centering
    \includegraphics[width=0.9\linewidth]{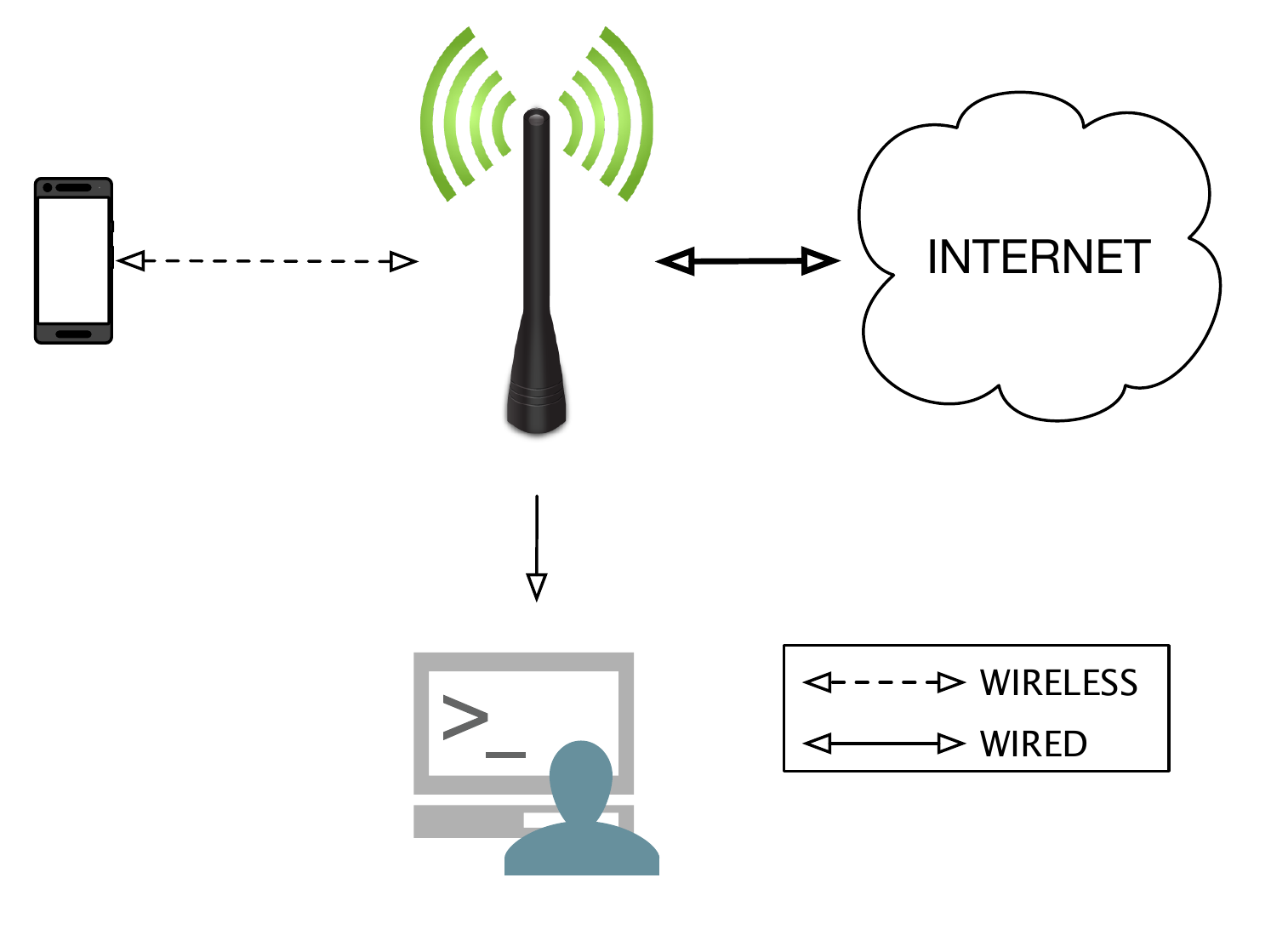}
  \caption{The attacker is able to passively capture the traffic from the target smartphone and the Internet access point (either Wi-Fi AP, or cellular base station).}
  \Description{The attacker is able to passively capture the traffic from the target smartphone and the Internet access point (either Wi-Fi AP, or cellular base station).}
  \label{fig:threat-model}   
\end{figure}

In our threat model an \emph{attacker} wants to deanonimize the apps on a \emph{target smartphone} that uses Tor. 
That is, he/she wants to recognize which apps are being used by the target smartphone at any given time.
We assume that the target is connected to the Internet through a wireless access point, either via a Wi-Fi LAN or via the cellular WAN, and that the attacker is able to passively capture the traffic between the target and the access point (see Figure~\ref{fig:threat-model}).
For example, the attacker could be the controller of the Wi-Fi Access Point or the Base Station the target is connected to.
We assume that the Tor client (i.e., an Onion Proxy) is installed in the smartphone itself and all apps' traffic passes through the Tor client.

%% file: methodology.tex
\section{Deanonymization Methodology}\label{sec:methodology}

\begin{figure}[t]
  \centering
  \includegraphics[width=\linewidth]{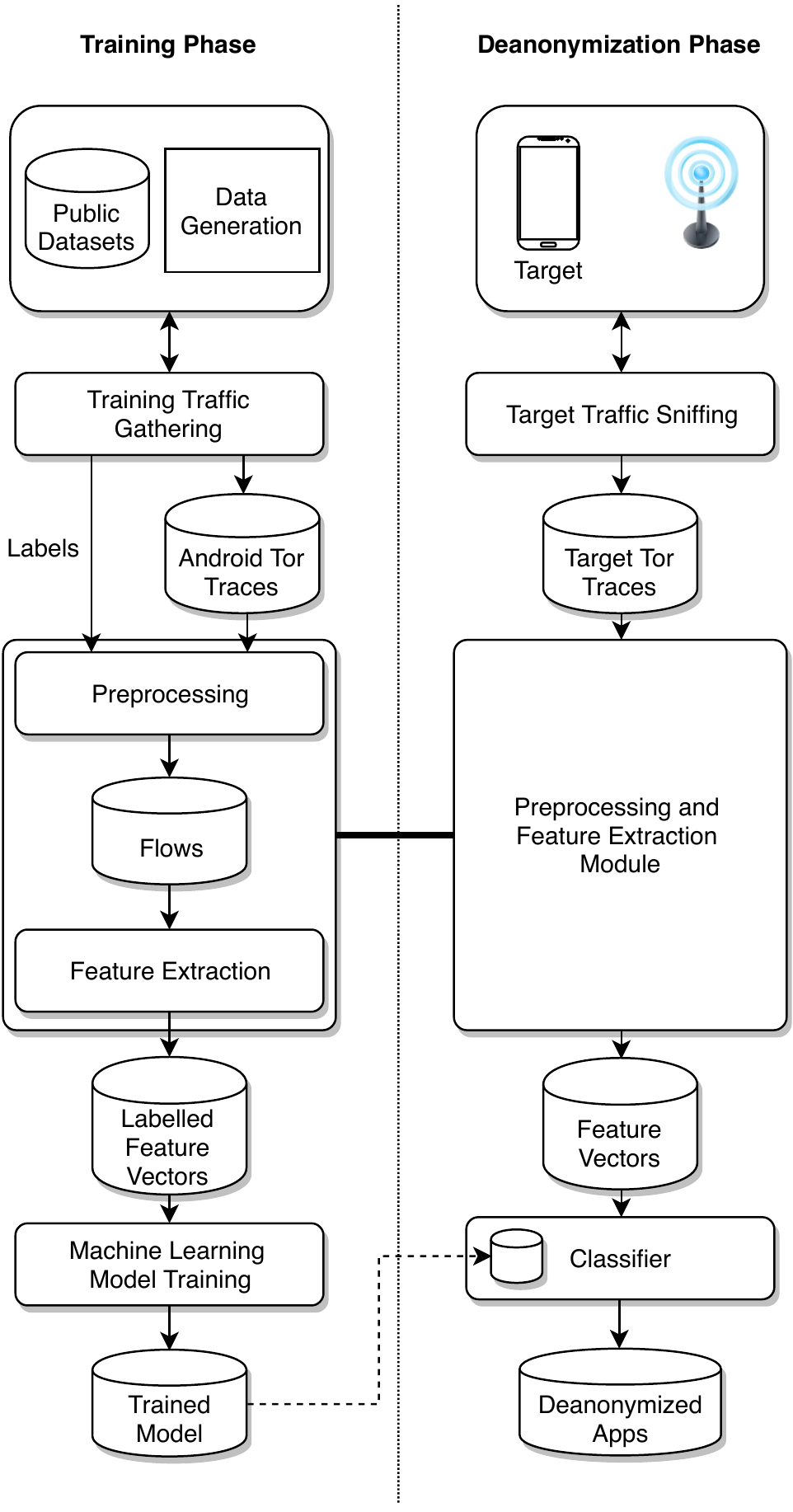}
  \caption{Overview of the deanonymization methodology.}\label{fig:methodology-overview}
  \Description{Overview of the deanonymization methodology.}
\end{figure}

Figure~\ref{fig:methodology-overview} shows an overview of our methodology for deanonymizing Android apps behaind Tor.
The assumption at the basis of the methodology is that different apps produce different network traffic patterns, which are discernible, through proper network traffic analysis, even when the traffic is anonymized through Tor. 

The methodology rely on a machine learning based network traffic analysis and consists of two distinct phases:
\begin{itemize}
\item \emph{Training Phase}: during which we build a machine learning model of the distinctive characteristics of apps' Tor traffic. This is the preparation phase of the attack.
\item \emph{Deanonymization Phase}: during which we conduct the actual attack against the target, by monitoring the target's traffic and using the model built in the previous phase to recognize which apps the victim is using. 
\end{itemize} 

In the next sections, we will describe each phase into details.

\subsection{Training Phase}
During the training phase we build a machine learning model of how different apps produce Tor traffic.
We assume that the attacker is interested in recognizing a predefined set of apps $\ClassSet = \{app_1,\dots,app_n\}$.
If the target is using an app which is not included in $\ClassSet$, our methodology will not be able to recognize that app.

The left part of Figure~\ref{fig:methodology-overview} shows the logical blocks of the training phase.
In the following sections we describe each logical block in details. 

\subsubsection{Training Traffic Gathering}
Since we assume a supervised learning process the first step is collecting a training dataset.
In particular, the very first step requires to gather, for each app in $\ClassSet$, raw Tor traffic traces (e.g. PCAP files~\cite{tcpdump}).
These traces can be picked from public datasets, if available (such as the one that we made available with this work), or can be generated synthetically, as described in section~\ref{sec:dataset-generation}.
The output of this step is a set of network traffic traces for each app in $\ClassSet$.

\subsubsection{Preprocessing and Feature Extraction Module}\label{sec:PFEM}
This module processes the network traces gathered at the previous step and extract the features that will be fed to the machine learning training algorithm.

\paragraph{Preprocessing}
For each network trace, we sort all TCP sessions (note that Tor only supports TCP) and we split sessions into flows.
A \emph{flow} is a portion of a TCP session of a predefined fixed duration $\FlowTimeout$, the \emph{flow timeout}.
We split each TCP sessions into flows of $\FlowTimeout$ seconds. When, we find a TCP packet with the FIN flag set, we stop splitting. Thus, the last flow of each TCP connection may actually last less than $\FlowTimeout$ seconds. 
The flow timeout is a configurable parameter of our methodology that has an impact on the deanonymization accuracy.
As detailed later, in section~\ref{sec:experiments}, we performed experiments with $\FlowTimeout = 10$ and $\FlowTimeout = 15$. 
The experiments with $\FlowTimeout = 10$ yielded slightly better results.

Once we have split all traces into flows, we label each flow with the app in $\ClassSet$ that has generated the corresponding traffic.
Therefore, the output of the preprocessing step is a labeled dataset of flows.

\paragraph{Feature Extraction}

For each flow $x_i$ we compute a vector of features $v_i = (f_1(x_i),\dots,f_m(x_i))$ and we assign it the corresponding label $app_j \in \ClassSet$.
Section~\ref{sec:features} reports the set of features that we employed in our Proof-of-Work.
The general methodology does not rely on a particular set of features. However, as always, the choice of such set strongly impacts on accuracy.
Our set of features has been derived from an experimental analysis involving various feature sets.

Since many machine learning algorithms (e.g., SVM and $k-$NN) work best with standardized features, for each component $y_{i,k} = f_k(x_i)$ (of each feature vector) we compute its standard score as:
\[
	z_{i,k} = \frac{y_{i,k} - \mu_k}{\sigma_k}
\]
where, $\mu_k$ and $\sigma_k$ are, respectively, the mean and standard deviation of feature $f_k$ computed over the dataset. 
The output is the training set, which is the dataset of labeled standardized feature vectors.
 
\subsubsection{Machine Learning Model Training}
During this step we feed the machine learning training algorithm with the training set build in the previous step.
The output is the trained model.
Our methodology does not rely on a particular machine learning model, but assumes a generic multi-class classifier whose set of classes is the set of apps $\ClassSet$.
In our experiments we tested three different classifiers based on, respectively, Random Forest, $k-$Nearest Neighbors and SVM.

\subsection{Deanonymization Phase}

In the deanonymization phase the actual attack against the target is performed.
The deanonymization phase consists of two logically distinct stages: a \emph{monitoring stage}, during which we sniff the target's network traffic, and a \emph{classification stage}, during which process the traces and output the classification.
In our Proof-of-Concept we adopt an \emph{offline} approach. That is, the two stages are not concurrent, they are performed subsequently. We first perform monitoring, collecting enough traces, and then we perform the classification.
However, our methodology is general enough to allow for an \emph{online} implementation, in which the two stages are actually executed simultaneously, and a new processing and classification step is performed as soon as the corresponding data is available.

The logical blocks of the deanonymization phase are shown in the right part of Figure~\ref{fig:methodology-overview}.
In the following sections we discuss each logical block in details. 

\subsubsection{Target Traffic Sniffing}
The first step of the deanonymization methodology requires the attacker to passively capture the target's network traffic. 
Our threat model assumes that the attacker has the means to do so.
For example, the attacker may be the controller of the wireless access point the target is connected to.
The output of this stage consists of the raw network traces.

\subsubsection{Preprocessing and Feature Extraction Module}
This module is exactly the same as the one already described in the training stage section (see~\ref{sec:PFEM}), with the only exception that in this case the traces and thus the corresponding feature vectors are not labeled.
Also the $\mu_k$ and $\sigma_k$ parameters used to standardize the features are those computed during the training phase (i.e., on the training set).

Thus, this module processes the traces coming from the previous step and outputs the corresponding (unlabeled) standardized feature vectors.

\subsubsection{Classifier}
In this step, each feature vector coming from the previous step is directly fed to the classifier that has been trained during the training stage.
For each feature vector the classifier outputs a class, namely one of the apps in $\ClassSet$.
The output of the classifier is also the output of the methodology, i.e., the deanonymized apps.

%% file: proof_of_concept.tex
\section{Proof-of-Concept}\label{sec:proof-of-concept}
This section presents our Proof-of-Concept implementing the methodology described in the previous section.
In the next sections we will describe the setup and the main logical blocks of the Proof-of-Concept. 
Then, we will discuss their implementation.
Finally, we will describe the set of features that we employ and the datasets that we built.

\subsection{Setup}\label{sec:setup}

\begin{figure}[t]
  \centering
  \includegraphics[width=\linewidth]{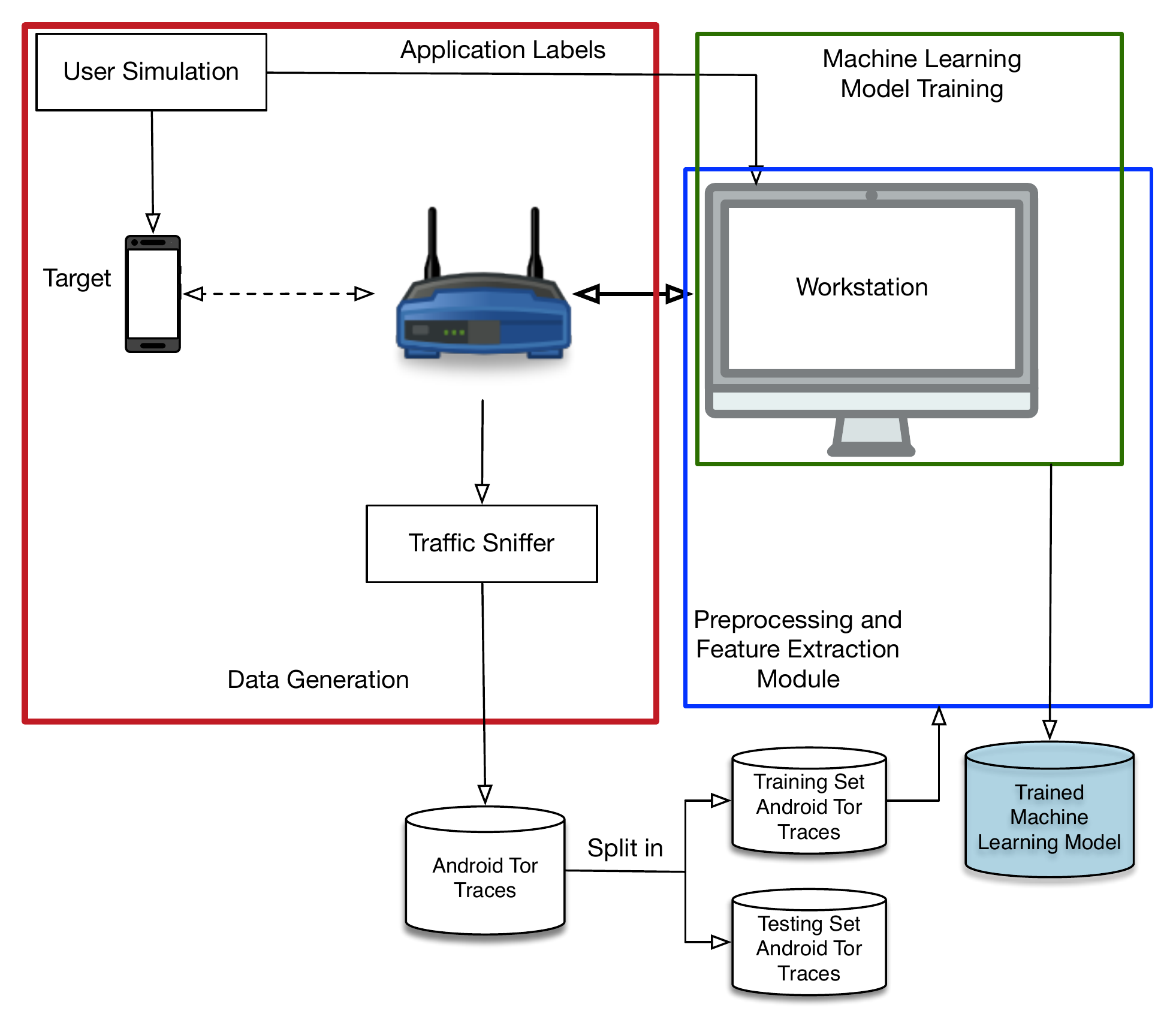}
  \caption{Proof-of-Concept setup.}\label{fig:poc-setup}
  \Description{Setup of the Proof-of-Concept.}
\end{figure}

For our proof-of-concept we built the setup shown in Figure~\ref{fig:poc-setup}.
The target smartphone connects to the Internet through a Wireless router.
The router is connected to a workstation. 
The latter is used to configure the router to capture the traffic of the target, and to run the preprocessing, feature extraction and machine learning tasks.

The router that we employed is a Xiaomi MiWiFi Nano running LEDE~\cite{lede}.
We performed experiments with two target smartphones: a Motorola
Moto G running its stock version of Android 6.0 and a Samsung Galaxy Nexus running LineageOS a mobile OS based on Android 6.0.
On both smartphones we use Orbot~\cite{orbot}, a proxy app that allows to use Tor on Android.
The workstation is a standard PC.

In the next sections we will discuss the main logical blocks of the Proof-of-Concept.

\subsubsection{Dataset Generation}\label{sec:dataset-generation}
Since, to the best of our knowledge, there are no publicly available datasets collecting Android Tor's traces, we generated our own datasets.
The basis of the dataset generation is the \emph{User Simulation} block shown in Figure~\ref{fig:poc-setup}.
For each app in $\ClassSet$, this block starts the app on the target smarphone and exercises it by simulating a user interaction with the app, so that it starts to generate a synthetic, yet as realistic as possible, network trace.
The router captures the traffic and stores it in a set of files through the \emph{Traffic Sniffer} module.
This dataset is then split into a training set and a test set.

\subsubsection{Processing and Feature Extraction Module}
The Processing and Feature Extraction module of our methodology is implemented in the workstation.
During the training stage it processes the training set and assign the labels registered by the User Simulation module to feature vectors, as shown in Figure~\ref{fig:poc-setup}.
During the deanonymization phase, the testing set is processed by the same module.

\subsubsection{Machine Learning Module}
The machine learning module is installed in the workstation. 
During the training stage it processes the training set of labeled feature vectors and outputs the trained model, as shown in Figure~\ref{fig:poc-setup}.
During the deanonymization phase, it uses the trained model to classify each feature vectors.

We tested three different classifiers, using Random Forest, $k-$Nearest Neighbor and SVM.
We performed the Stratified $k$-Fold Cross Validation~\cite{Kohavi:1995:SCB:1643031.1643047} during training.

\subsection{Implementation}\label{sec:implementation}

In this section we discuss the implementation of the modules described in section~\ref{sec:setup}.

\subsubsection{User Simulation}
The User Simulation module has been implemented through AndroidViewClient~\cite{androidviewclient} a python library that helps scripting Android apps GUI exercisers.
Moreover, we employed Culebra GUI and CulebraTester~\cite{culebra}. 
These tools record the actions performed by a real user on an Android device and convert them in a script to reproduce them.
Culebra GUI can produce python scripts that are compatible with AndroidViewClient.
We employed these tools mainly to identify the correct UI elements to interact with, for each app.
Then, we elaborated different simulation scripts for each app, depending on their typical use.
We reported the details about simulation of the various apps in appendix~\ref{sec:user-simulation}.

\subsubsection{Traffic Sniffer}
The traffic sniffer is implemented by running Tcpdump~\cite{tcpdump} on the router, to capture network traffic, and Wireshark~\cite{wireshark} on the workstation for simple network analysis.

\subsubsection{Preprocessing and Feature Extraction Module}
The preprocessing and feature extraction module has been implemented from scratch as a set of python scripts processing the traces PCAP files, splitting the TCP sessions into flows and extracting the features on them.

\subsubsection{Machine Learning module}
The machine learning module has been implemented through Scikit-Learn~\cite{scikit-learn}, a well-known python library that implements many machine learning algorithms.

\subsection{Features}\label{sec:features}
In our Proof-of-Concept we employed 3 types of features. 
In the following we describe all the features that we used.

\subsubsection{Time-based Features}
Since Tor's relay cells (those that transport the actual payload) are fixed sized, initially we concentrated on time-based, rather than size-based features.
Indeed, since Tor aims to be a low-latency network, it does not introduce random delays when routing packets.
Thus, time-based features appear to be a good candidate to characterize apps traffic over Tor.
In particular, we employed the following features, given that they led to good results in the context of recognition of traffic classes in desktop environments~\cite{habibi}:
\begin{itemize}
\item \emph{FIAT} (Forward Inter Arrival Time): time between two outgoing packets;
\item \emph{BIAT} (Backward Inter Arrival Time): time between two incoming packets;
\item \emph{FLOWIAT} (Flow Inter Arrival Time): time between two packets, no
matter the direction;
\item \emph{Active time}: amount of time a flow is active;
\item \emph{Idle time}: amount of time a flow is idle;
\item \emph{Flow bytes per second}: number of bytes per second;
\item \emph{Flow packets per second}: number of packets per second;
\item \emph{Duration}: duration of the flow in seconds.
\end{itemize}

For all the above features except the last three, we actually compute 4 statistical values: minimum, maximum, mean and standard deviation.
Moreover, the active and idle time depends on a configurable threshold, the \emph{activity timeout} $\ActivityTimeout$.
We performed experiments with $\ActivityTimeout = 2$ and $\ActivityTimeout = 5$ seconds. 

\subsubsection{Packet Direction and Burst Features}
Packet direction and burst features have also been proven to be effective in the context of website fingerprinting on desktop environments~\cite{waknn}.
Packet direction indicates whether a packet is going forward, from the source (the Tor client) to the destination, or backward, i.e., in the opposite direction.
A burst instead is an uninterrupted sequence of packets in the same direction.
After a preliminary visual analysis, with the Wireshark~\cite{wireshark} tool, we observed distinctive patterns, in terms of bursts, in the network traces of the various apps.
Therefore we decided to enrich our feature set with the following features:
\begin{itemize}
\item Direction of the first 10 packets (of the flow);
\item Incoming Bursts: number of bursts, bursts mean length, length of the longest burst;
\item Outgoing Bursts: number of bursts, bursts mean length, length of the longest burst;
\item Lengths of the first 10 incoming bursts;
\item Lengths of the first 10 outgoing bursts. 
\end{itemize}

\subsubsection{Size-based Features}
Event though relay cells are fixed sized, Tor uses variable-length cells primarily used for traffic control.
As a preliminary analysis, we counted the number of packets for each packet size and we observed that, while there is a high variability of packet sizes, there is a relatively small set of packet sizes that accounts for most of the traffic.
Thus, we decided to introduce a feature for each of the ten most frequent packet sizes.
These, were (in order of higher frequency) $1500$, $595$, $583$, $2960$, $1097$, $1384$, $151$, $1126$, $1109$ and $233$ bytes.
We soon decided to discard size $2960$, as this exceeds the MTU ($1500$ byes) and thus represents a reassembled packet.
Each feature is a counter of the number of packets of that size observed in the flow. 

\subsection{Datasets}\label{sec:datasets}
We collected two datasets of network traces:
\begin{itemize}
\item \emph{Reduced Connection Padding Dataset}: $11.24$ GB of traces  collected with the reduced connection padding activated (see section~\ref{sec:reducedpadding}), which is the default configuration on Orbot.
\item \emph{Full Connection Padding Dataset}: $9.84$ GB of traces collected with the (full) connection padding activated (see section~\ref{sec:fullpadding}).
\end{itemize}

Both datasets consists of a set of PCAP files~\cite{tcpdump} containing the raw captured packets.
In both datasets we collected about 4 hours of network traffic for each of the following apps: Dailimotion, Facebook, Instagram, Replaio Radio, Skype, Spotify, TorBrowser Alpha, Twitch, uTorrent, YouTube.

%% file: experiments.tex
\section{Experimental Evaluation}\label{sec:exp-eval}
This section reports the experimental evaluation that we carried out to asses the accuracy of the methodology in deanonymizing apps.
In section~\ref{sec:experiments} we introduce the experiments that we performed. In section~\ref{sec:eval-method} we discuss the evaluation methodology. 
Finally, in section~\ref{sec:results} we present and comment the obtained results.

\subsection{Experiments}\label{sec:experiments}
We performed several experiments using the prototype implementation of our methodology described in the Proof-of-Concept section (see section~\ref{sec:proof-of-concept}).

For each experiment we vary the following settings:
\begin{itemize}
\item \emph{Tor's connection padding}: \textit{Reduced} or \textit{Full}, depending on whether we use the dataset with reduced connection padding or full connection padding (see section~\ref{sec:datasets});
\item \emph{Flow Timeout} ($\FlowTimeout$): either 10 or 15 seconds (see section~\ref{sec:PFEM});
\item \emph{Activity Timeout} ($\ActivityTimeout$): either 2 or 5 seconds (see section~\ref{sec:features});
\item \emph{Presence of the Web Browser app}: Yes/No.
\end{itemize}
In particular, the last setting indicates whether the traces related to the usage of the web browser app are included in the experiment's dataset or not.
The choice of performing experiments for both cases is motivated by the fact that, according to our experiments, the web browser app seems to be the most difficult to recognize among those that we tested, and significantly reduces the accuracy of the methodology.
Therefore, we show the accuracy that can be achieved in both cases, by including or by ignoring such app.

Moreover, in each experiment we tested the performance of three different classifiers: Random Forest~\cite{randomforest}, $k-$Nearest Neighbors ($k$-NN)~\cite{knn} and C-SVM classifier (SVC)~\cite{cristianini2000introduction}.

\begin{table}[t]
  \caption{The complete set of performed experiments (Flow Timeout and Activity Timeout are in seconds).}
  \label{tab:experiments}
  \begin{tabular}{lcccc}
    \toprule
    \multirowcell{2}{Experiment} & \multirowcell{2}{Connection\\ Padding} & \multirowcell{2}{Flow\\ Timeout} & \multirowcell{2}{Activity\\ Timeout} & \multirowcell{2}{Web\\ Browser}\\
    & & & & \\
    \midrule
    Experiment 1 & Reduced & 10 & 2 & Yes \\
    Experiment 2 & Reduced & 10 & 2 & No \\
    Experiment 3 & Full & 10 & 2 & Yes \\
    Experiment 4 & Full & 10 & 2 & No \\
    Experiment 5 & Reduced & 10 & 5 & Yes \\
    Experiment 6 & Reduced & 10 & 5 & No \\
    Experiment 7 & Full & 10 & 5 & Yes \\
    Experiment 8 & Full & 10 & 5 & No \\
    Experiment 9 & Reduced & 15 & 2 & Yes \\
    Experiment 10 & Reduced & 15 & 2 & No \\
    Experiment 11 & Full & 15 & 2 & Yes \\
    Experiment 12 & Full & 15 & 2 & No \\
    Experiment 13 & Reduced & 15 & 5 & Yes \\
    Experiment 14 & Reduced & 15 & 5 & No \\
    Experiment 15 & Full & 15 & 5 & Yes \\
    Experiment 16 & Full & 15 & 5 & No \\
  \bottomrule
\end{tabular}
\end{table}

Table~\ref{tab:experiments} summarizes the complete set of experiments that we performed and their settings.

\subsection{Evaluation Methodology}\label{sec:eval-method}
For each experiment we evaluate the performance achieved by our Proof-of-Concept, namely the performance of the classifier.
We asses both the overall performance of the classifier and the performance achieved on a per-class basis, so as to highlight whether some apps are more easily recognized than others. 

The per-class performance are computed in terms of precision, recall, F1 score and accuracy computed for each class in $\ClassSet$.
The formulas employed to compute such metrics are reported in Table~\ref{tab:per-class-perf-measures}.

\begin{table}
\caption{Per-class performance metrics, where $TP_i, TN_i, FP_i, FN_i$ are, respectively, the number of true positives, true negatives, false positives and false negatives of class $i$.}
\label{tab:per-class-perf-measures}
\begin{tabular}{lll}
\toprule
Measure & Formula \\ 
\midrule
\multirow{2}{*}{Precision$_i$} & \multirow{2}{*}{$\frac{TP_{i}}{TP_{i}+FP_{i}}$} \\
& & \\
\multirow{2}{*}{Recall$_i$} & \multirow{2}{*}{$\frac{TP_{i}}{TP_{i}+FN_{i}}$} \\
& &\\ 
\multirow{2}{*}{F1 Score$_i$} & \multirow{2}{*}{$2\frac{Precision_i \cdot Recall_i}{Precision_i + Recall_i}$} \\
& &\\ 
\multirow{2}{*}{Accuracy$_i$} & \multirow{2}{*}{$\frac{TP_{i}+TN_{i}}{TP_{i}+FP_{i}+TN_{i}+FN_{i}}$} \\
& & \\ 
 \bottomrule
\end{tabular}
\end{table}

The overall classifier performance are computed by averaging the per-class metrics.
Note that precision, recall and F1 score are averaged according to two criteria: micro and macro.
The two criteria account differently for imbalances in the dataset (i.e., uneven proportion of samples per classes).
The micro criteria biases the corresponding metrics towards the most populated classes, while the macro criteria treats all classes equally~\cite{performanceevaluation}.
The formulas employed to compute such metrics are reported in Table~\ref{tab:overall-perf-measures}.
Note, that micro precision and micro recall (and thus micro F1 score), are mathematically equivalent.
Thus, when presenting the results of the experiments we will only report the micro F1 score.

\begin{table}
\caption{Overall classifier performance measures, where $TP_i, TN_i, FP_i, FN_i$ are, respectively, the number of true positives, true negatives, false positives and false negatives of class $i$, and $n$ is the total number of classes.}
\label{tab:overall-perf-measures}
\begin{tabular}{lll}
\toprule
Measure & Formula \\ 
\midrule
\multirow{2}{*}{Micro Precision} & \multirow{2}{*}{$\frac{\sum_i TP_{i}}{\sum_i \left(TP_{i}+FP_{i}\right)}$} \\
  & & \\
\multirow{2}{*}{Micro Recall} & \multirow{2}{*}{$\frac{\sum_i TP_{i}}{\sum_i \left(TP_{i}+FN_{i}\right)}$} \\
& &\\ 
\multirow{2}{*}{Micro F1 Score} & \multirow{2}{*}{$2\frac{Micro Precision \cdot Micro Recall}{Micro Precision + Micro Recall}$} \\
& &\\ 
\multirow{2}{*}{Macro Precision} & \multirow{2}{*}{$\frac{\sum_i \frac{TP_{i}}{TP_{i}+FP_{i}}}{n} = \frac{\sum_i Precision_i}{n}$} \\
& & \\
\multirow{2}{*}{Macro Recall} & \multirow{2}{*}{$\frac{\sum_i \frac{TP_{i}}{TP_{i}+FN_{i}}}{n} = \frac{\sum_i Recall_i}{n}$} \\
& &\\ 
\multirow{2}{*}{Macro F1 Score} & \multirow{2}{*}{$2\frac{Macro Precision \cdot Macro Recall}{Macro Precision + Macro Recall}$} \\
& &\\ 
\multirow{2}{*}{Average Accuracy} & \multirow{2}{*}{$\frac{ \sum_i\frac{TP_{i}+TN_{i}}{TP_{i}+FP_{i}+TN_{i}+FN_{i}}}{n} = \frac{\sum_i Accuracy_i}{n}$} \\
& & \\ 
\multirow{2}{*}{Error Rate} & \multirow{2}{*}{$\frac{ \sum_i\frac{FP_{i}+FN_{i}}{TP_{i}+FP_{i}+TN_{i}+FN_{i}}}{n}$} \\
& & \\ 
 \bottomrule
\end{tabular}
\end{table}

Formulas in Tables~\ref{tab:overall-perf-measures} and \ref{tab:per-class-perf-measures} are defined in terms of the number of true positives, false positives, true negatives and false negatives.
These are defined as follows.
Let $\widehat{v_i}$ be the standardized feature vector of flow $x_i$.
Let $c \in \ClassSet$ be the actual class of $\hat{v_i}$, that is, the one corresponding to the app that generated the traffic of flow $x_i$. 
Let $\overline{c} \in \ClassSet$ be the app that is output by the classifier.
Then:
\begin{itemize}
\item $TP_i$: number of feature vectors for which $\overline{c} = c = app_i$;
\item $FP_i$: number of feature vectors for which $\overline{c} = app_i$ and $c \ne app_i$;
\item $TN_i$: number of feature vectors for which $\overline{c} \ne app_i$ and $c \ne app_i$;
\item $FN_i$: number of feature vectors for which $\overline{c} \ne app_i$ and $c = app_i$;
\end{itemize}

\subsection{Results}\label{sec:results}
In this section we present the results of the experimental evaluation. 
As already stated, for ease of presentation, we only discuss the experiments corresponding to the combination of the Flow and Activity Timeout parameters that yielded the best results. 
Namely, experiments 1, 2, 3 and 4. The results of the other experiments are reported for completeness in the appendix~\ref{sec:all_results}.
In the following sections we present and comment the results of experiments 1-4, while in section~\ref{sec:results-summary} we summarizes the overall results and draw the conclusions of the experimental evaluation. 

\subsubsection{Experiment 1}

\begin{figure}[t]
  \centering
  \includegraphics[width=\linewidth]{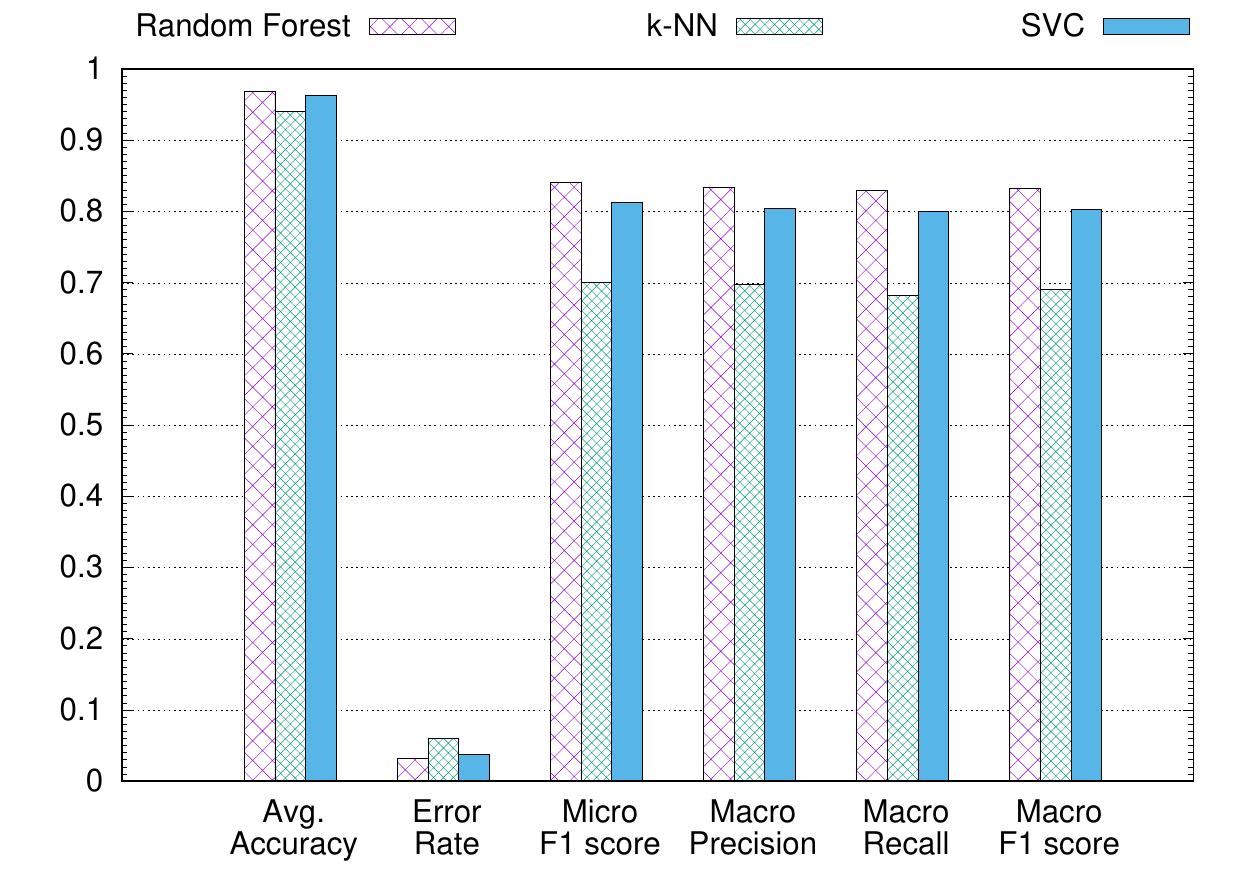}
  \caption{Classifiers' overall performance for Experiment~1.}\label{fig:overall-res-exp1}
  \Description{Classifiers' overall performance for Experiment 1.}
\end{figure}

This experiment is performed on the Reduced Connection Padding Dataset, with $\FlowTimeout = 10$ and $\ActivityTimeout = 2$, and includes the Tor Browser app.
Figure~\ref{fig:overall-res-exp1} shows the overall performance of the three classifiers.
We achieved the best performance with the Random Forest classifier (with respect to all metrics).
In particular, Random Forest achieves an accuracy of $0.968$ and an error rate of $0.032$ with a (macro) F1 score of $0.83$.
With $k$-NN we obtained the worst results, while SVC performs slightly worse than Random Forest. 

\begin{table*}[t]
\centering
\caption{Per-class performance of each classifier for Experiment 1.}
\label{tab:per-class-res-exp1}
\begin{tabular}{lcccccccccccc}
 & \multicolumn{4}{c}{\textbf{Random Forest}} & \multicolumn{4}{c}{\textbf{$k-$NN}} & \multicolumn{4}{c}{\textbf{SVC}} \\ \hline
\multicolumn{1}{c|}{\textbf{APP}} & \textbf{PR.} & \textbf{REC.} & \textbf{F1} & \multicolumn{1}{c|}{\textbf{ACC.}} & \textbf{PR.} & \textbf{REC.} & \textbf{F1} & \multicolumn{1}{c|}{\textbf{ACC.}} & \textbf{PR.} & \textbf{REC.} & \textbf{F1} & \textbf{ACC.} \\ \hline
\multicolumn{1}{l|}{dailymotion} & 0.83 & 0.77 & 0.8 & \multicolumn{1}{c|}{0.96} & 0.56 & 0.58 & 0.57 & \multicolumn{1}{c|}{0.91} & 0.74 & 0.72 & 0.73 & 0.95 \\
\multicolumn{1}{l|}{facebook} & 0.9 & 0.84 & 0.87 & \multicolumn{1}{c|}{0.98} & 0.62 & 0.7 & 0.66 & \multicolumn{1}{c|}{0.94} & 0.86 & 0.85 & 0.86 & 0.97 \\
\multicolumn{1}{l|}{instagram} & 0.79 & 0.86 & 0.82 & \multicolumn{1}{c|}{0.94} & 0.58 & 0.67 & 0.62 & \multicolumn{1}{c|}{0.88} & 0.77 & 0.83 & 0.8 & 0.94 \\
\multicolumn{1}{l|}{replaio\_radio} & 0.99 & 0.98 & 0.98 & \multicolumn{1}{c|}{1.0} & 0.98 & 0.96 & 0.97 & \multicolumn{1}{c|}{0.99} & 0.98 & 0.98 & 0.98 & 0.99 \\
\multicolumn{1}{l|}{skype} & 0.99 & 0.96 & 0.97 & \multicolumn{1}{c|}{1.0} & 0.97 & 0.94 & 0.95 & \multicolumn{1}{c|}{0.99} & 0.98 & 0.95 & 0.97 & 0.99 \\
\multicolumn{1}{l|}{spotify} & 0.67 & 0.65 & 0.66 & \multicolumn{1}{c|}{0.94} & 0.56 & 0.48 & 0.52 & \multicolumn{1}{c|}{0.92} & 0.63 & 0.66 & 0.65 & 0.93 \\
\multicolumn{1}{l|}{torbrowser\_alpha} & 0.68 & 0.77 & 0.72 & \multicolumn{1}{c|}{0.97} & 0.6 & 0.47 & 0.53 & \multicolumn{1}{c|}{0.95} & 0.67 & 0.71 & 0.69 & 0.96 \\
\multicolumn{1}{l|}{twitch} & 0.83 & 0.87 & 0.85 & \multicolumn{1}{c|}{0.97} & 0.68 & 0.76 & 0.71 & \multicolumn{1}{c|}{0.93} & 0.83 & 0.83 & 0.83 & 0.96 \\
\multicolumn{1}{l|}{utorrent} & 0.9 & 0.91 & 0.9 & \multicolumn{1}{c|}{0.98} & 0.82 & 0.69 & 0.75 & \multicolumn{1}{c|}{0.96} & 0.85 & 0.84 & 0.85 & 0.97 \\
\multicolumn{1}{l|}{youtube} & 0.76 & 0.69 & 0.72 & \multicolumn{1}{c|}{0.95} & 0.61 & 0.57 & 0.59 & \multicolumn{1}{c|}{0.93} & 0.72 & 0.63 & 0.67 & 0.95 \\
\hline
\end{tabular}
\end{table*}

Table~\ref{tab:per-class-res-exp1} reports the per-class performance metrics.
For all classifiers, we observe a certain variability in how accurate the classifier is in recognizing the various apps. 
Looking at the F1 score, Spotify, Tor Browser and YouTube appear to be the most difficult apps to recognize.
Indeed, by looking directly at the data, we observed that these three apps are often confused, one for another.
Since both Spotify and YouTube provide streaming contents, they probably generate too similar traffic patterns sometimes, that mislead the classifiers.
The same reasoning probably applies to Tor Browser.
Indeed, when visiting websites providing streaming contents, this app may be confused with a app such as YouTube or Spotify.
Indeed, webpages may embed streaming content, including YouTube videos themselves.

\subsubsection{Experiment 2}

\begin{figure}[t]
  \centering
  \includegraphics[width=\linewidth]{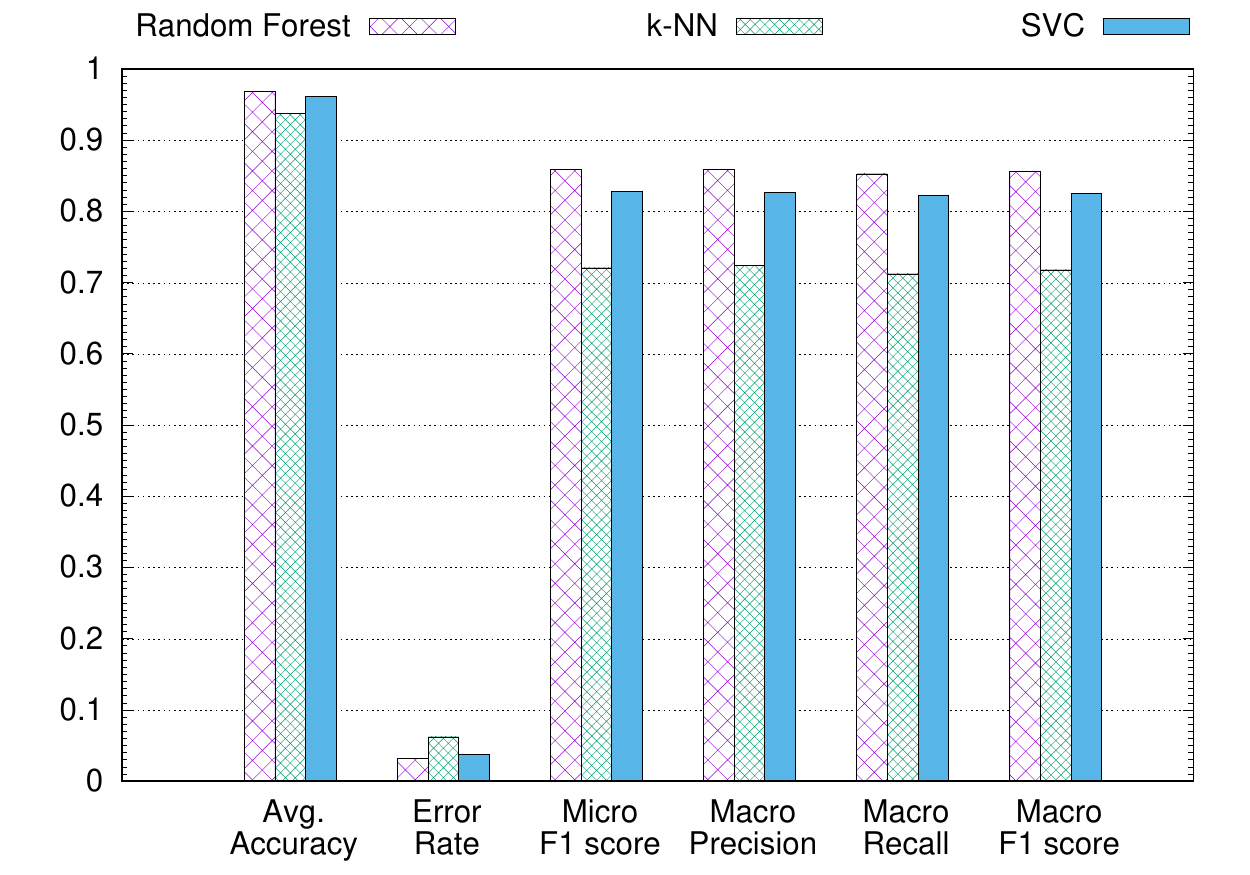}
  \caption{Classifiers' overall performance for Experiment 2.}\label{fig:overall-res-exp2}
  \Description{Classifiers' overall performance for Experiment 2.}
\end{figure}

This experiment is performed on the same dataset as Experiment~1 (Reduced Connection Padding Dataset), with the same parameters ($\FlowTimeout = 10$ and $\ActivityTimeout = 2$), but does not include the Tor Browser app.
Figure~\ref{fig:overall-res-exp2} shows the overall performance of the three classifiers.
Same as Experiment~1, the best performance have been obtained with the Random Forest classifier.
This classifier achieved $0.969$ accuracy, with an error rate of $0.031$ and a (macro) F1 score of $0.855$.
Also in this case $k$-NN performed the worst, while SVC achieved better results.
In this experiment we obtained better overall performance than in Experiment~1.
This is due to the fact that, as already pointed out, the Tor Browser app is often misclassified due to its variability in usage patterns.
Thus, by not considering it all classifiers make less mistakes.

\begin{table*}[t]
\centering
\caption{Per-class performance of each classifier for Experiment 2.}
\label{tab:per-class-res-exp2}
\begin{tabular}{lcccccccccccc}
 & \multicolumn{4}{c}{\textbf{Random Forest}} & \multicolumn{4}{c}{\textbf{$k-$NN}} & \multicolumn{4}{c}{\textbf{SVC}} \\ \hline
\multicolumn{1}{c|}{\textbf{APP}} & \textbf{PR.} & \textbf{REC.} & \textbf{F1} & \multicolumn{1}{c|}{\textbf{ACC.}} & \textbf{PR.} & \textbf{REC.} & \textbf{F1} & \multicolumn{1}{c|}{\textbf{ACC.}} & \textbf{PR.} & \textbf{REC.} & \textbf{F1} & \textbf{ACC.} \\ \hline
\multicolumn{1}{l|}{dailymotion} & 0.83 & 0.78 & 0.8 & \multicolumn{1}{c|}{0.96} & 0.56 & 0.58 & 0.57 & \multicolumn{1}{c|}{0.9} & 0.74 & 0.72 & 0.73 & 0.94 \\
\multicolumn{1}{l|}{facebook} & 0.9 & 0.84 & 0.87 & \multicolumn{1}{c|}{0.98} & 0.65 & 0.71 & 0.68 & \multicolumn{1}{c|}{0.94} & 0.86 & 0.85 & 0.85 & 0.97 \\
\multicolumn{1}{l|}{instagram} & 0.79 & 0.87 & 0.83 & \multicolumn{1}{c|}{0.94} & 0.6 & 0.67 & 0.63 & \multicolumn{1}{c|}{0.88} & 0.77 & 0.82 & 0.79 & 0.93 \\
\multicolumn{1}{l|}{replaio\_radio} & 0.99 & 0.98 & 0.99 & \multicolumn{1}{c|}{1.0} & 0.98 & 0.96 & 0.97 & \multicolumn{1}{c|}{0.99} & 0.98 & 0.98 & 0.98 & 0.99 \\
\multicolumn{1}{l|}{skype} & 0.99 & 0.96 & 0.98 & \multicolumn{1}{c|}{1.0} & 0.98 & 0.95 & 0.96 & \multicolumn{1}{c|}{0.99} & 0.98 & 0.96 & 0.97 & 0.99 \\
\multicolumn{1}{l|}{spotify} & 0.72 & 0.74 & 0.73 & \multicolumn{1}{c|}{0.95} & 0.6 & 0.5 & 0.55 & \multicolumn{1}{c|}{0.92} & 0.67 & 0.72 & 0.69 & 0.94 \\
\multicolumn{1}{l|}{twitch} & 0.84 & 0.87 & 0.85 & \multicolumn{1}{c|}{0.97} & 0.68 & 0.76 & 0.72 & \multicolumn{1}{c|}{0.93} & 0.84 & 0.83 & 0.83 & 0.96 \\
\multicolumn{1}{l|}{utorrent} & 0.9 & 0.93 & 0.91 & \multicolumn{1}{c|}{0.98} & 0.83 & 0.71 & 0.77 & \multicolumn{1}{c|}{0.96} & 0.87 & 0.87 & 0.87 & 0.97 \\
\multicolumn{1}{l|}{youtube} & 0.77 & 0.71 & 0.74 & \multicolumn{1}{c|}{0.95} & 0.63 & 0.56 & 0.59 & \multicolumn{1}{c|}{0.93} & 0.73 & 0.66 & 0.69 & 0.95 \\
\hline
\end{tabular}
\end{table*}

The per-class performance metrics are reported Table~\ref{tab:per-class-res-exp3}.
Looking at the F1 score, we observe that, similarly to Experiment~1, the apps for which we achieve the worst performance are Spotify and YouTube.
This, again, is due to the fact that they are often confused one for the other.
However, with respect to Experiment~1 all metrics increases for both apps.
This is due to the fact that they were often misclassified as Tor Browser.

\subsubsection{Experiment 3}

\begin{figure}[t]
  \centering
  \includegraphics[width=\linewidth]{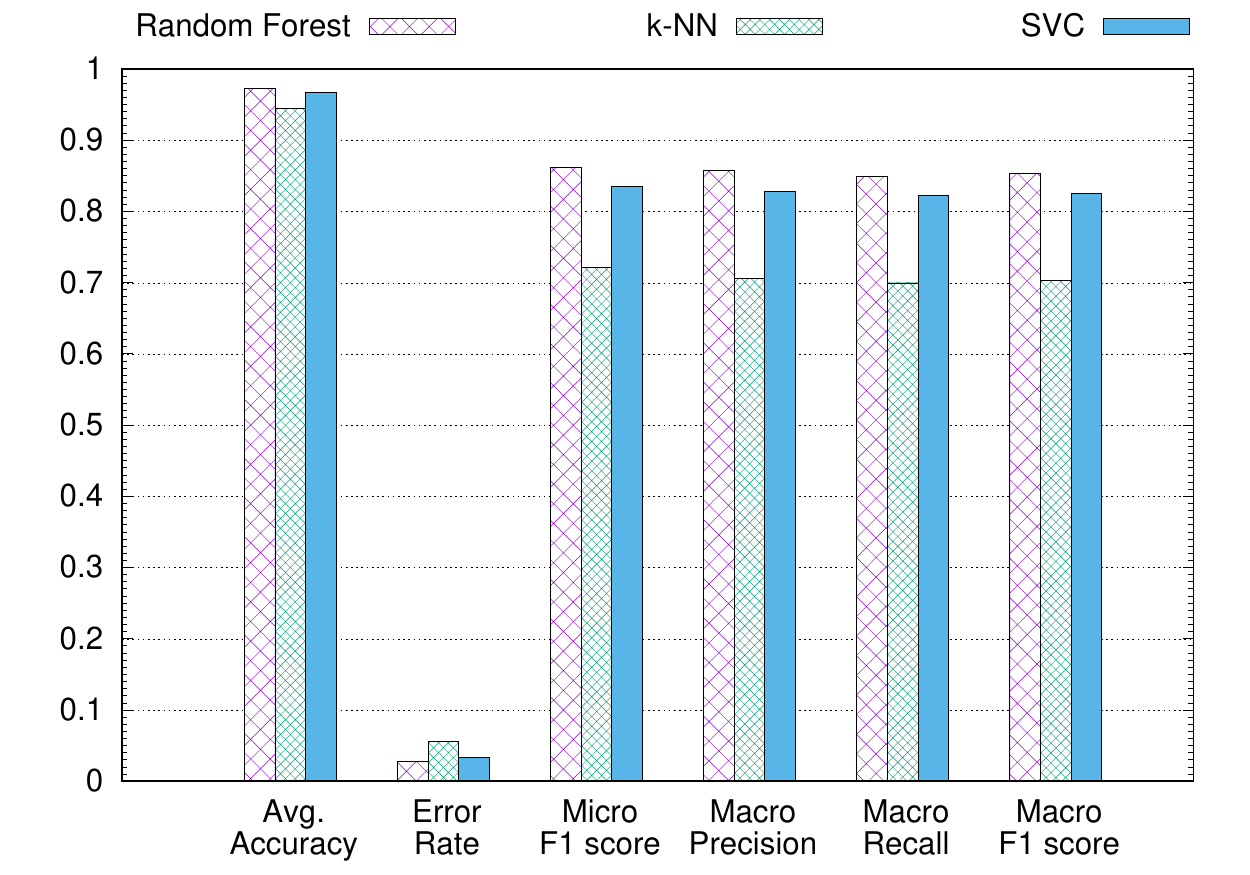}
  \caption{Classifiers' overall performance for Experiment 3.}\label{fig:overall-res-exp3}
  \Description{Classifiers' overall performance for Experiment 3.}
\end{figure}

This experiment is performed on the Full Connection Padding Dataset, with $\FlowTimeout = 10$ and $\ActivityTimeout = 2$, and includes the Tor Browser app.
As in the previous two experiments, Random Forest gave the best results, followed by SVC and $k$-NN.
The accuracy achieved through Random Forest is $0.972$, with an error rate of $0.028$ and a (macro) F1 score of $0.853$.

\begin{table*}[t]
\centering
\caption{Per-class performance of each classifier for Experiment 3.}
\label{tab:per-class-res-exp3}
\begin{tabular}{lcccccccccccc}
 & \multicolumn{4}{c}{\textbf{Random Forest}} & \multicolumn{4}{c}{\textbf{$k-$NN}} & \multicolumn{4}{c}{\textbf{SVC}} \\ \hline
\multicolumn{1}{c|}{\textbf{APP}} & \textbf{PR.} & \textbf{REC.} & \textbf{F1} & \multicolumn{1}{c|}{\textbf{ACC.}} & \textbf{PR.} & \textbf{REC.} & \textbf{F1} & \multicolumn{1}{c|}{\textbf{ACC.}} & \textbf{PR.} & \textbf{REC.} & \textbf{F1} & \textbf{ACC.} \\ \hline
\multicolumn{1}{l|}{dailymotion} & 0.87 & 0.81 & 0.84 & \multicolumn{1}{c|}{0.97} & 0.66 & 0.72 & 0.69 & \multicolumn{1}{c|}{0.95} & 0.8 & 0.77 & 0.78 & 0.97 \\
\multicolumn{1}{l|}{facebook} & 0.78 & 0.72 & 0.75 & \multicolumn{1}{c|}{0.96} & 0.48 & 0.52 & 0.5 & \multicolumn{1}{c|}{0.91} & 0.71 & 0.72 & 0.72 & 0.95 \\
\multicolumn{1}{l|}{instagram} & 0.72 & 0.7 & 0.71 & \multicolumn{1}{c|}{0.94} & 0.47 & 0.47 & 0.47 & \multicolumn{1}{c|}{0.89} & 0.67 & 0.7 & 0.68 & 0.93 \\
\multicolumn{1}{l|}{replaio\_radio} & 0.95 & 0.97 & 0.96 & \multicolumn{1}{c|}{0.99} & 0.88 & 0.94 & 0.91 & \multicolumn{1}{c|}{0.98} & 0.96 & 0.96 & 0.96 & 0.99 \\
\multicolumn{1}{l|}{skype} & 0.98 & 0.95 & 0.97 & \multicolumn{1}{c|}{0.99} & 0.93 & 0.92 & 0.93 & \multicolumn{1}{c|}{0.98} & 0.98 & 0.94 & 0.96 & 0.99 \\
\multicolumn{1}{l|}{spotify} & 0.82 & 0.84 & 0.83 & \multicolumn{1}{c|}{0.96} & 0.69 & 0.68 & 0.68 & \multicolumn{1}{c|}{0.93} & 0.76 & 0.78 & 0.77 & 0.95 \\
\multicolumn{1}{l|}{torbrowser\_alpha} & 0.81 & 0.69 & 0.75 & \multicolumn{1}{c|}{0.97} & 0.64 & 0.35 & 0.45 & \multicolumn{1}{c|}{0.94} & 0.78 & 0.67 & 0.72 & 0.97 \\
\multicolumn{1}{l|}{twitch} & 0.86 & 0.92 & 0.89 & \multicolumn{1}{c|}{0.98} & 0.68 & 0.77 & 0.72 & \multicolumn{1}{c|}{0.94} & 0.86 & 0.87 & 0.87 & 0.97 \\
\multicolumn{1}{l|}{utorrent} & 0.99 & 0.99 & 0.99 & \multicolumn{1}{c|}{1.0} & 0.94 & 0.98 & 0.96 & \multicolumn{1}{c|}{0.99} & 0.98 & 0.99 & 0.98 & 1.0 \\
\multicolumn{1}{l|}{youtube} & 0.79 & 0.88 & 0.83 & \multicolumn{1}{c|}{0.96} & 0.68 & 0.65 & 0.66 & \multicolumn{1}{c|}{0.92} & 0.79 & 0.82 & 0.8 & 0.95 \\
\hline
\end{tabular}
\end{table*}

Table~\ref{tab:per-class-res-exp3} summarizes the per-class results.
Differently from Experiments~1 and 2, by looking at the F1 score, we observe that the apps that mislead the classifiers the most are Facebook, Instagram and Tor Browser.
This is not surprising.
Indeed, if we think of the typical usage patterns of the apps that we considered in our experiments, Facebook, Instagram and Tor Browser are the ones with the highest idle periods (the user ``think time''), as opposed to the other apps, that mainly provide streaming content (typically, with less frequent and shorter idle periods).
Since the connection padding mechanism is activated by idle periods, it is normal to observe a performance degradation when using full connection padding rather than the reduced one.

\subsubsection{Experiment 4}

\begin{figure}[t]
  \centering
  \includegraphics[width=\linewidth]{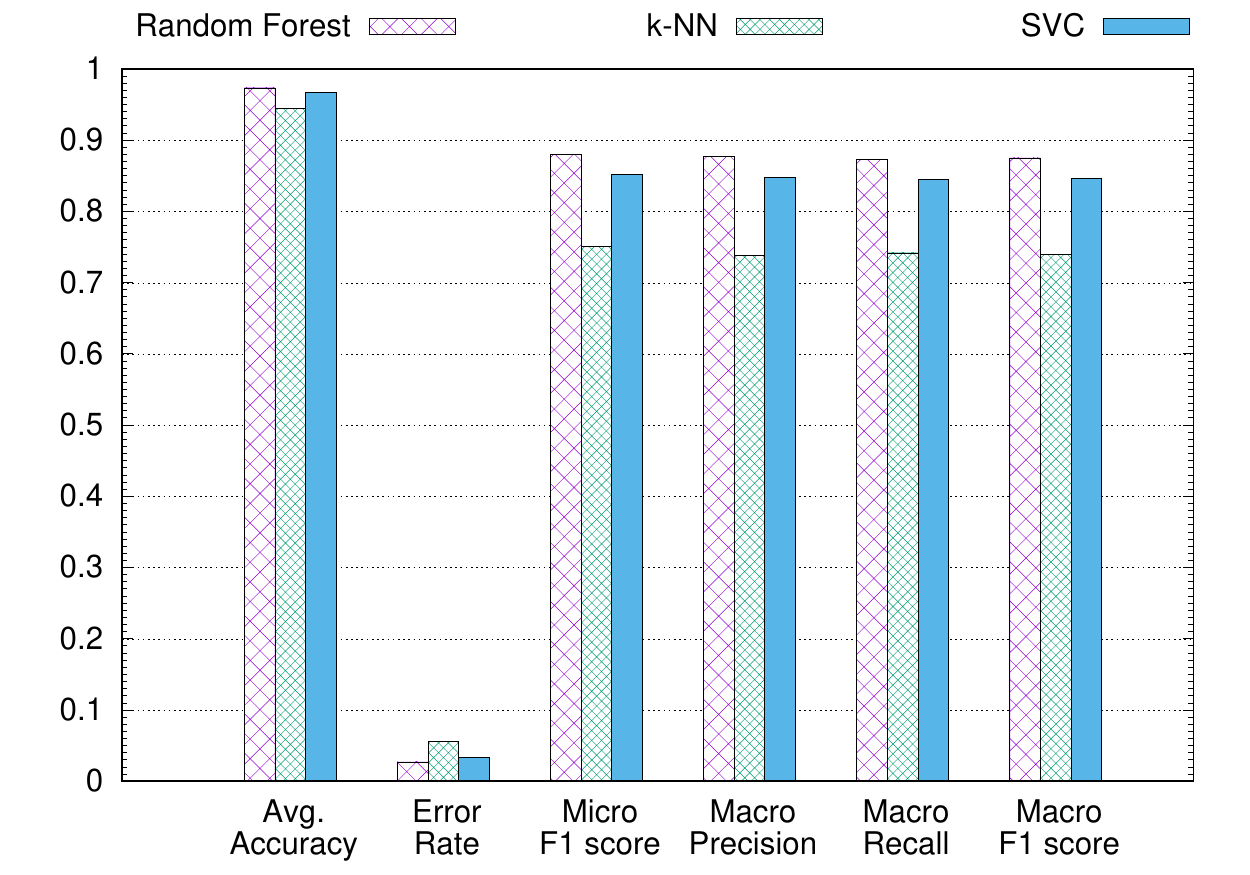}
  \caption{Classifiers' overall performance for Experiment 4.}\label{fig:overall-res-exp4}
  \Description{Classifiers' overall performance for Experiment 4.}
\end{figure}

This experiment is performed on the same dataset as Experiment~3 (Full Connection Padding Dataset), with the same parameters ($\FlowTimeout = 10$ and $\ActivityTimeout = 2$), but does not include the Tor Browser app.
As in all other experiments, Random Forest provided the best results, while $k$-NN performed the worst.
With Random Forest we obtained an accuracy of $0.973$, an error rate of $0.0268$ and a (macro) F1 score of $0.875$.

\begin{table*}[t]
\centering
\caption{Per-class performance of each classifier for Experiment 4.}
\label{tab:per-class-res-exp4}
\begin{tabular}{lcccccccccccc}
 & \multicolumn{4}{c}{\textbf{Random Forest}} & \multicolumn{4}{c}{\textbf{$k-$NN}} & \multicolumn{4}{c}{\textbf{SVC}} \\ \hline
\multicolumn{1}{c|}{\textbf{APP}} & \textbf{PR.} & \textbf{REC.} & \textbf{F1} & \multicolumn{1}{c|}{\textbf{ACC.}} & \textbf{PR.} & \textbf{REC.} & \textbf{F1} & \multicolumn{1}{c|}{\textbf{ACC.}} & \textbf{PR.} & \textbf{REC.} & \textbf{F1} & \textbf{ACC.} \\ \hline
\multicolumn{1}{l|}{dailymotion} & 0.88 & 0.82 & 0.85 & \multicolumn{1}{c|}{0.97} & 0.67 & 0.71 & 0.69 & \multicolumn{1}{c|}{0.94} & 0.81 & 0.77 & 0.79 & 0.96 \\
\multicolumn{1}{l|}{facebook} & 0.82 & 0.73 & 0.77 & \multicolumn{1}{c|}{0.96} & 0.51 & 0.52 & 0.52 & \multicolumn{1}{c|}{0.91} & 0.73 & 0.73 & 0.73 & 0.95 \\
\multicolumn{1}{l|}{instagram} & 0.75 & 0.71 & 0.73 & \multicolumn{1}{c|}{0.94} & 0.51 & 0.49 & 0.5 & \multicolumn{1}{c|}{0.89} & 0.69 & 0.7 & 0.69 & 0.93 \\
\multicolumn{1}{l|}{replaio\_radio} & 0.95 & 0.97 & 0.96 & \multicolumn{1}{c|}{0.99} & 0.88 & 0.94 & 0.91 & \multicolumn{1}{c|}{0.98} & 0.96 & 0.96 & 0.96 & 0.99 \\
\multicolumn{1}{l|}{skype} & 0.98 & 0.96 & 0.97 & \multicolumn{1}{c|}{0.99} & 0.96 & 0.92 & 0.94 & \multicolumn{1}{c|}{0.98} & 0.97 & 0.95 & 0.96 & 0.99 \\
\multicolumn{1}{l|}{spotify} & 0.84 & 0.86 & 0.85 & \multicolumn{1}{c|}{0.97} & 0.73 & 0.69 & 0.71 & \multicolumn{1}{c|}{0.94} & 0.8 & 0.81 & 0.8 & 0.95 \\
\multicolumn{1}{l|}{twitch} & 0.87 & 0.92 & 0.9 & \multicolumn{1}{c|}{0.98} & 0.71 & 0.77 & 0.74 & \multicolumn{1}{c|}{0.94} & 0.88 & 0.87 & 0.87 & 0.97 \\
\multicolumn{1}{l|}{utorrent} & 0.99 & 0.99 & 0.99 & \multicolumn{1}{c|}{1.0} & 0.96 & 0.98 & 0.97 & \multicolumn{1}{c|}{0.99} & 0.98 & 0.99 & 0.99 & 1.0 \\
\multicolumn{1}{l|}{youtube} & 0.81 & 0.88 & 0.84 & \multicolumn{1}{c|}{0.96} & 0.71 & 0.65 & 0.68 & \multicolumn{1}{c|}{0.92} & 0.8 & 0.83 & 0.81 & 0.95 \\
\hline
\end{tabular}
\end{table*}

Table~\ref{tab:per-class-res-exp3} summarizes the per-class results.
The same considerations made for Experiment~3 holds for Experiment~4.
That is, looking at the F1 score, Facebook and Instagram are the apps that hinders all classifiers the most.
Since we excluded the Tor Browser app, we notice a general improvement of the performance.
This, again, is due to the fact that this app is particularly difficult to discern from other apps.

\subsubsection{Results Summary}\label{sec:results-summary}

\begin{table*}
  \caption{Summary of the results of Experiments 1-4.}
  \label{tab:results-summary}
  \begin{tabular}{lcccccc}
    \toprule
    Experiment & Avg. Accuracy & Error Rate & Micro F1 score & Macro Precision & Macro Recall & Macro F1 score\\
    \midrule
    Experiment~1 & 0.968 & 0.032 & 0.840 & 0.834 & 0.830 & 0.832 \\
    
    Experiment~2 & 0.969 & 0.031 & 0.859 & 0.859 & 0.852 & 0.855 \\
    
    Experiment~3 & 0.972 & 0.028 & 0.861 & 0.857 & 0.849 & 0.853 \\
    
    Experiment~4 & 0.973 & 0.0268 & 0.880 & 0.877 & 0.872 & 0.875 \\
	\bottomrule
\end{tabular}
\end{table*}

In all experiments we achieved the best results with the Random Forest classifier.
Table~\ref{tab:results-summary} shows a comparison of the results obtained in each experiment through this classifier.
In all experiments we obtained comparable accuracy ($\sim 0.97$).
As expected, all performance metrics slightly improve when we do not consider the Tor Browser app. 
Indeed, the type of the visited website strongly impact on the characteristics of the generated traffic, which makes this app sometimes be confused with other apps.
For example, when visiting a webpage with streaming content the Tor Browser app might be confused with a streaming app (such as Spotify or YouTube).

A counterintuitive result that we obtained is that apparently the use of Tor's (full) connection padding actually improved the accuracy over the use reduced connection padding.
If we look at the per-class results (Tables~\ref{tab:per-class-res-exp1}-\ref{tab:per-class-res-exp4}) we notice that the performance on Facebook and Instagram apps actually worsen significantly.
Also the recall of the Tor Browser app worsen significantly, though its precision improves, which means that the proportion of false negatives increases (the app is more often confused with others), while the number of false positives decreases (other apps are less frequently confused with Tor Browser).
The fact that these three apps are more often misclassified when using full padding is what we expected.
Indeed, as already pointed out, their typical use patterns involve more frequent ``think times'' and, thus, idle periods, which trigger the connection padding mechanism.
On the other hand, the other apps are mainly characterized by a ``streaming'' pattern, thus involving extremely less frequent idle periods, which explains why for the majority of them the performance does not worsen.
However, it does not explain why they improve.
Clearly, the padding mechanism has a strong impact on the time-based features (see section~\ref{sec:features}), especially the active/idle time.
Our guess is that the full padding mechanism is actually activated statistically more often for some of these streaming apps and less often for others, which actually results in a better separation of the corresponding classes.
We plan to better investigate this aspect as future work. 
 

%% file: conclusion.tex
\section{Conclusion}\label{sec:conclusion}

In this work we have shown that Tor is vulnerable to app deanonymization.
We described a general methodology to perform an attack against a target smartphone which allows to unveil which app the victim is using.
The proposed methodology performs network traffic analysis based on a supervised machine learning approach. 
It leverages the fact that different apps produce different recognizable traffic patterns even when protected by Tor.
We also provided a Proof-of-Concept that implements the methodology, that we employed to assess the accuracy that it can achieve in deanonymizing apps.
We performed several experiments achieving an accuracy of $97.3\%$ and a F1 score of $87.5\%$. 
We made the software of the Proof-of-Concept, as well as the datasets that we built during the experiments, publicly available, so that it can be used to assess Tor's vulnerability to this attack, compare alternative methodologies and test possible countermeasures. 

As future work we plan to experiment with additional machine learning algorithms.
Moreover, in this work we adopted a multi-class classifier approach.
That is, we trained a single classifier on all possible classes.
We plan to extend our experimental evaluation by testing alternative binary-class approaches (such as \emph{one-vs-all} and \emph{one-vs-one}), in which we employ several binary classifiers in place of a single multi-class classifier.
Another improvement to this work may be to enlarge the datasets with a richer set of apps.

%% file: appendix.tex

\appendix

\section{User Simulation}\label{sec:user-simulation}

This section describes how we simulated the user interaction in our Proof-of-Work.

\subsection{Tor Browser}
The user activity on the Tor Browser app has been simulated through a python script that visits webpages randomly sampled from a list of the top 10,000 sites extracted from the Majestic Million dataset~\cite{majesticmillion}.
The script spend a randomly drawn amount of time on each webpage, before navigating to the next one.

\subsection{Instagram}
To simulate the user interaction with Instagram, we created a new account and added the Socialblade's top 500 most followed profiles~\cite{instagramlist}. 
The simulation script generates random swipe inputs on the Instagram app to scroll the main page up and down with random delays.
Swipe down inputs are generated with higher probability than swipe up inputs, as a user browsing Instagram posts would typically scroll the page from top to bottom.
After a random number of swipes there is a 30\% probability that the user decides to visit another random profile, or otherwise a 30\% probability that the user will push the like button on the current Instagram post.

\subsection{Facebook}
The simulation of the user interaction with the Facebook app is very similar to that of Instagram.
First we create a Facebook account for the user and we add a list of followed pages derived from Socialblade's top 500 most liked Facebook Pages~\cite{facebooklist}.
Similarly to that of Instagram, the simulation script scrolls the posts in the main page of the Facebook app, by generating random swipe inputs with random delays. After a random number of swipes there is a 30\% probability that the user pushes the like button on the post showing on the screen.

\subsection{Skype}
Skype calls have been generated by starting calls with an audio source near the smartphone microphone.

\subsection{uTorrent}
The uTorrent app is a Torrent client and, therefore, it does not require a complex user interaction. 
We simply add some torrent file to the app, and it starts the download.

\subsection{Dailymotion, Replaio Radio, Spotify, Twitch, YouTube}
Also this apps do not require a very complex interaction with the user.
We start each app on some streaming content and leave the app in execution.

\section{Other Experiments Results}\label{sec:all_results}

\begin{figure*}[h]
  \centering
  \begin{minipage}[c]{0.5\textwidth}
	  \includegraphics[width=\linewidth]{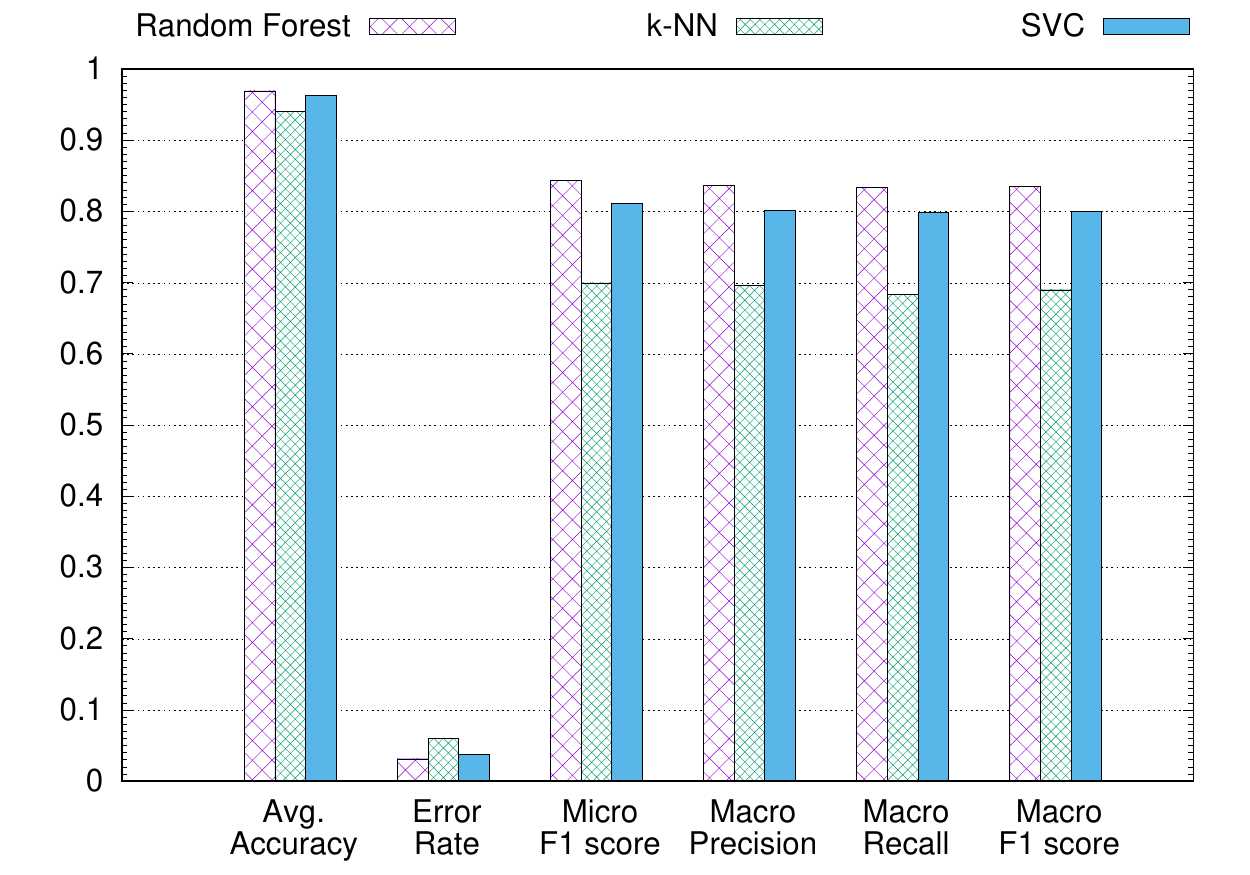}
	  \caption{Classifiers' overall performance for Experiment~5.}\label{fig:overall-res-exp5}
	  \Description{Classifiers' overall performance for Experiment 5.}
  \end{minipage}%
  \begin{minipage}[c]{0.5\textwidth}
      \includegraphics[width=\linewidth]{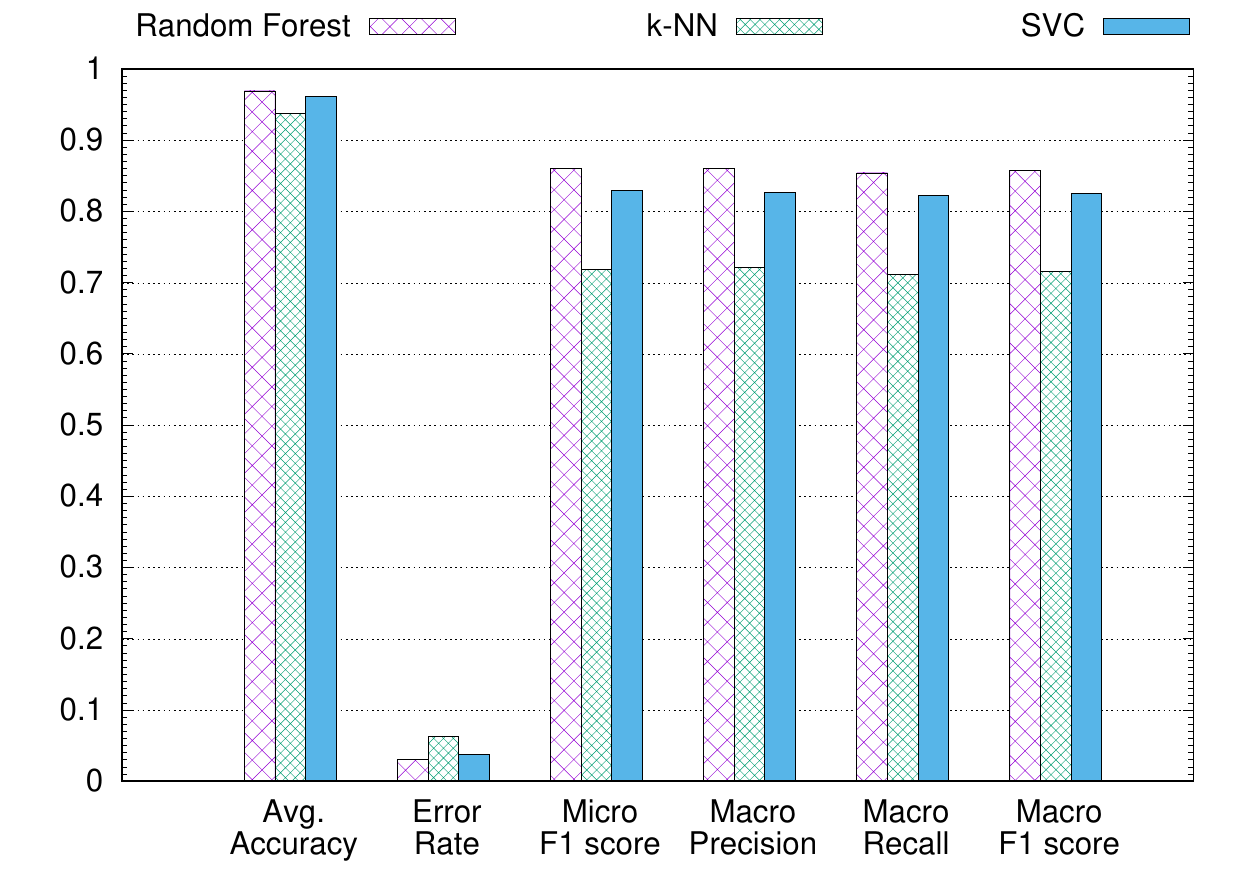}
	  \caption{Classifiers' overall performance for Experiment~6.}\label{fig:overall-res-exp6}
	  \Description{Classifiers' overall performance for Experiment 6.}
  \end{minipage}
  
  \begin{minipage}[c]{0.5\textwidth}
    \includegraphics[width=\linewidth]{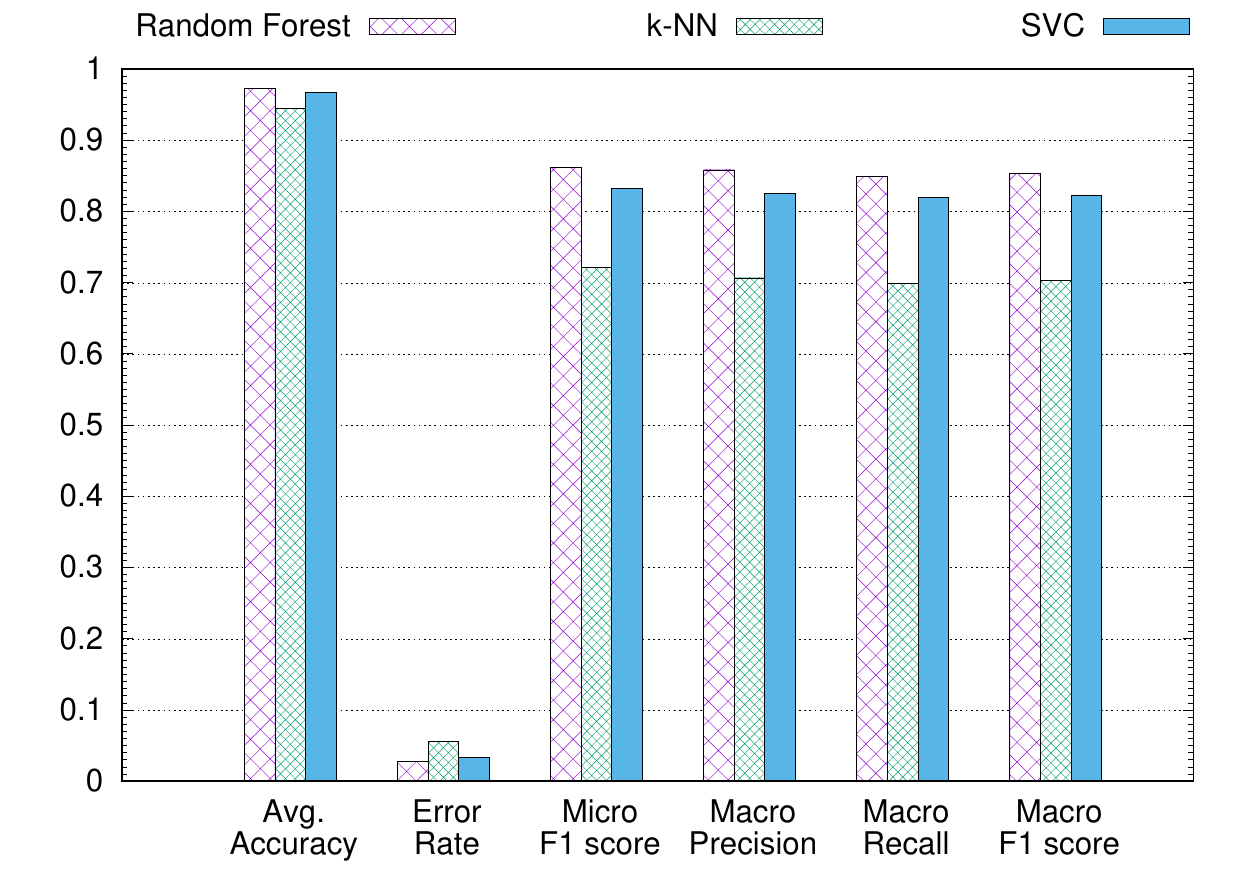}
    \caption{Classifiers' overall performance for Experiment~7.}\label{fig:overall-res-exp7}
    \Description{Classifiers' overall performance for Experiment 7.}
  \end{minipage}%
  \begin{minipage}[c]{0.5\textwidth}
    \includegraphics[width=\linewidth]{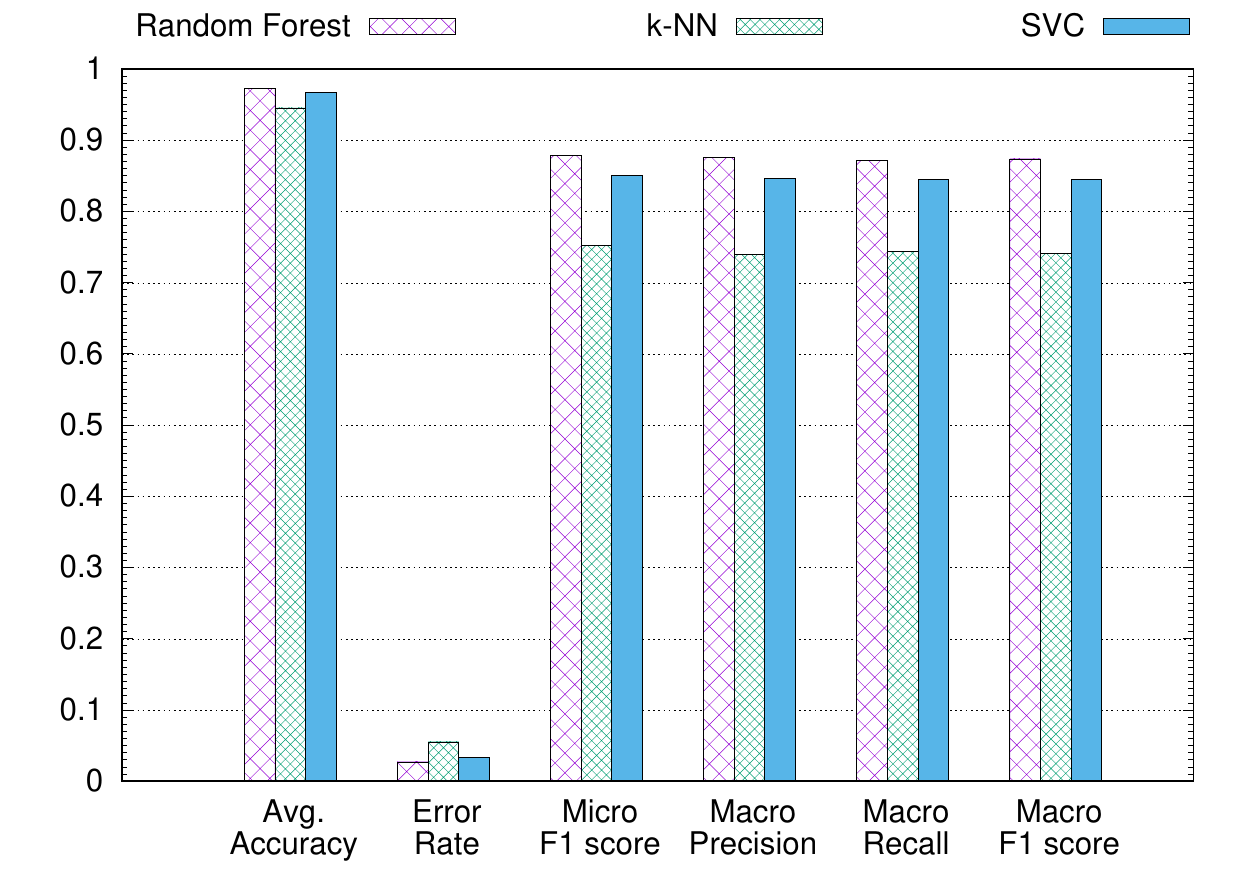}
    \caption{Classifiers' overall performance for Experiment~8.}\label{fig:overall-res-exp8}
    \Description{Classifiers' overall performance for Experiment 8.}
  \end{minipage}
  
  \begin{minipage}[c]{0.5\textwidth}
    \includegraphics[width=\linewidth]{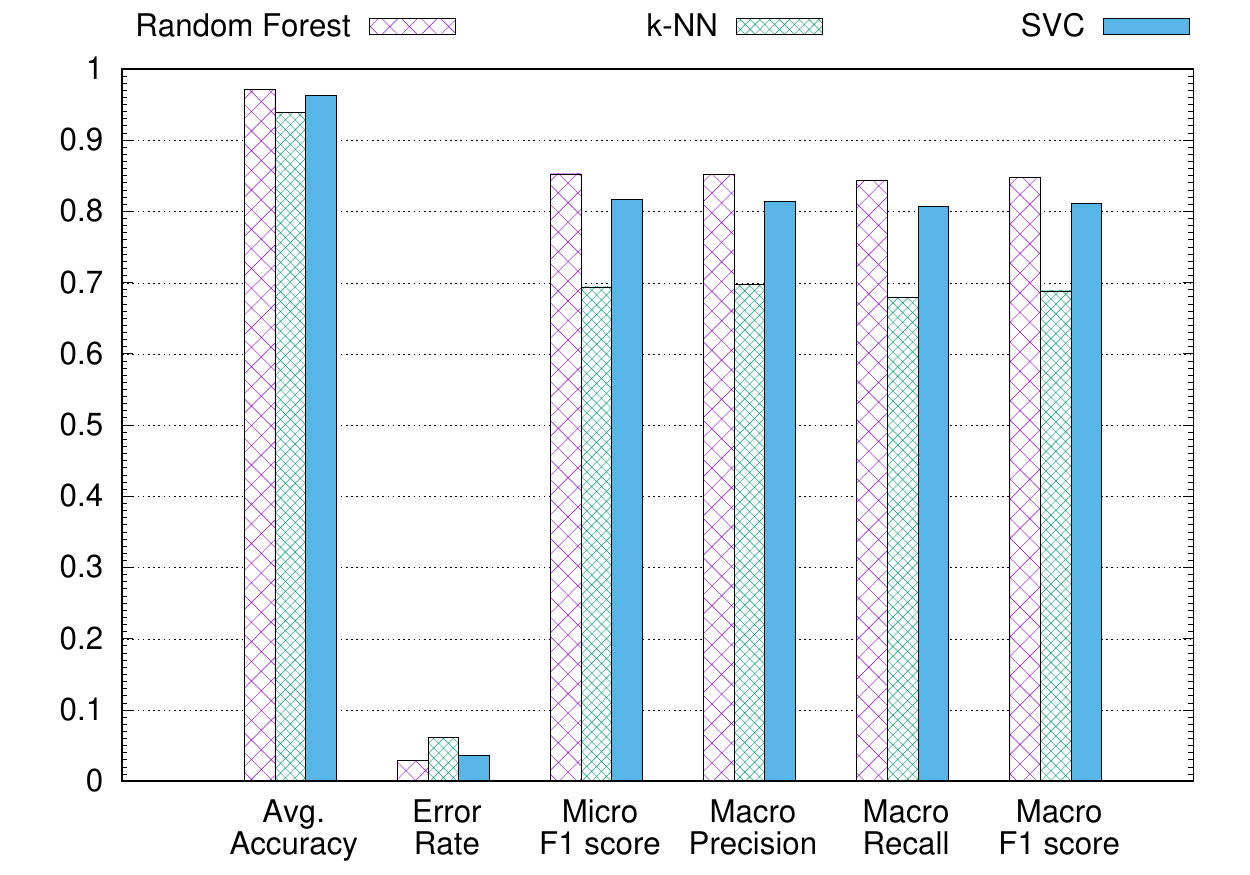}
    \caption{Classifiers' overall performance for Experiment~9.}\label{fig:overall-res-exp9}
    \Description{Classifiers' overall performance for Experiment 9.}
  \end{minipage}%
  \begin{minipage}[c]{0.5\textwidth}
    \includegraphics[width=\linewidth]{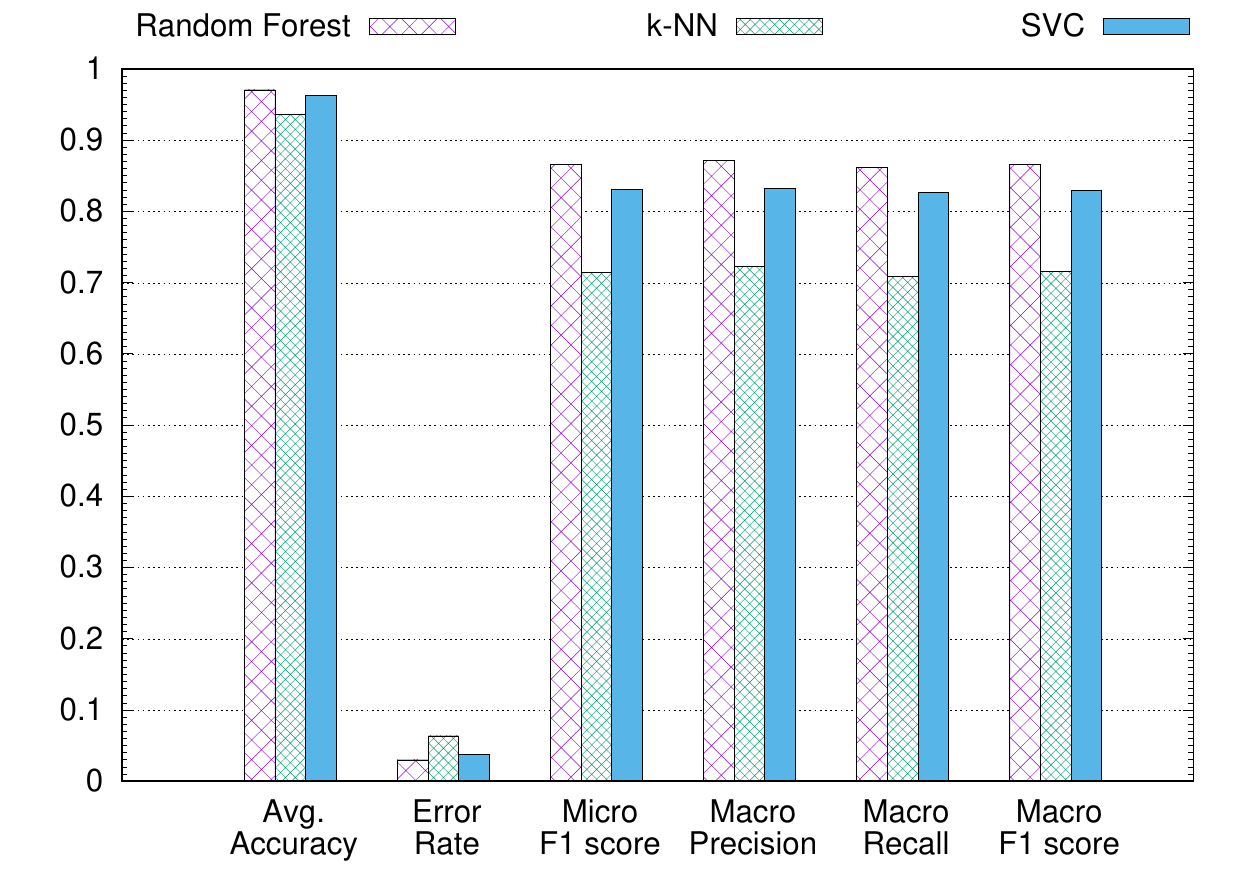}
    \caption{Classifiers' overall performance for Experiment~10.}\label{fig:overall-res-exp10}
    \Description{Classifiers' overall performance for Experiment 10.}
  \end{minipage}
\end{figure*}

\begin{figure*}[h]
  \centering
  \begin{minipage}[c]{0.5\textwidth}
	  \includegraphics[width=\linewidth]{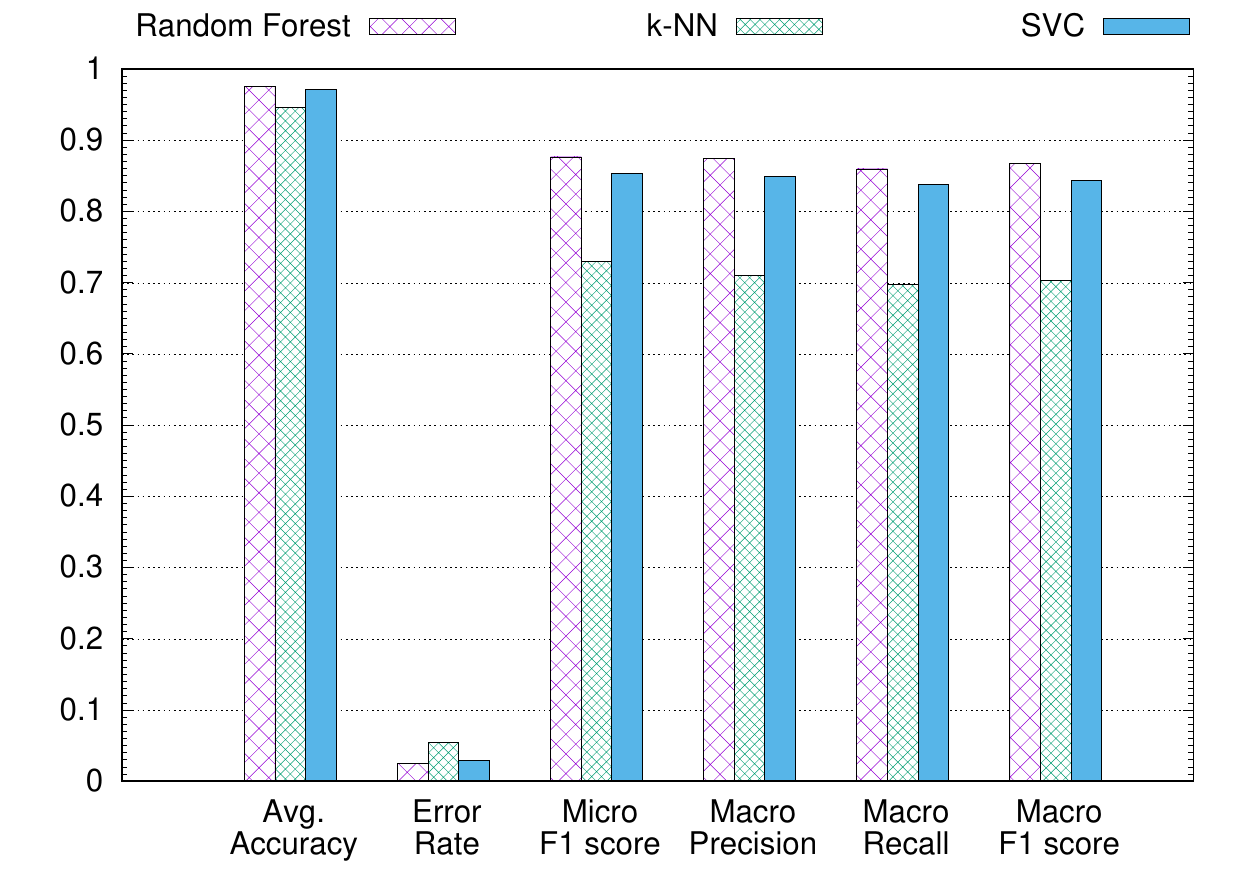}
	    \caption{Classifiers' overall performance for Experiment~11.}\label{fig:overall-res-exp11}
	    \Description{Classifiers' overall performance for Experiment 11.}
  \end{minipage}%
  \begin{minipage}[c]{0.5\textwidth}
      \includegraphics[width=\linewidth]{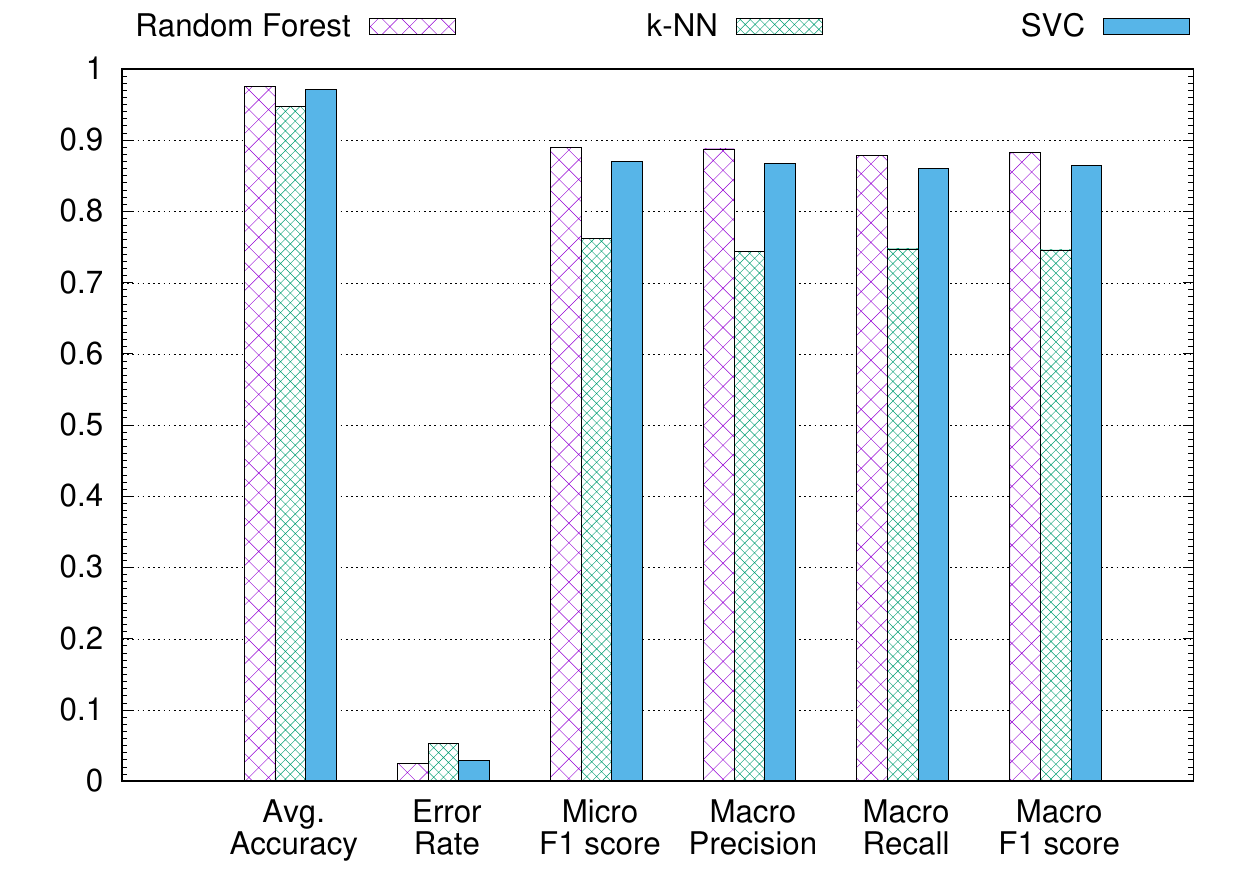}
        \caption{Classifiers' overall performance for Experiment~12.}\label{fig:overall-res-exp12}
        \Description{Classifiers' overall performance for Experiment 12.}
  \end{minipage}
  
  \begin{minipage}[c]{0.5\textwidth}
    \includegraphics[width=\linewidth]{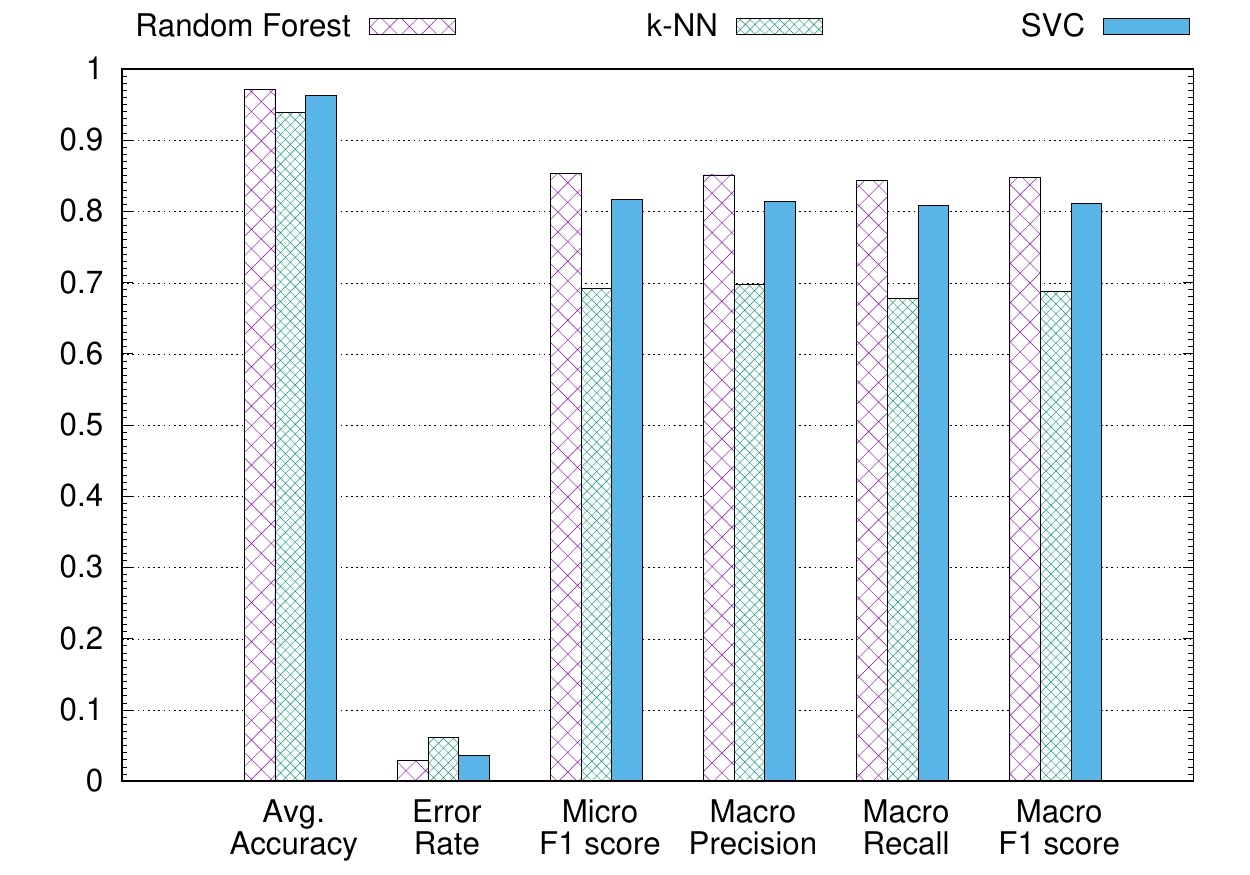}
      \caption{Classifiers' overall performance for Experiment~13.}\label{fig:overall-res-exp13}
      \Description{Classifiers' overall performance for Experiment 13.}
  \end{minipage}%
  \begin{minipage}[c]{0.5\textwidth}
    \includegraphics[width=\linewidth]{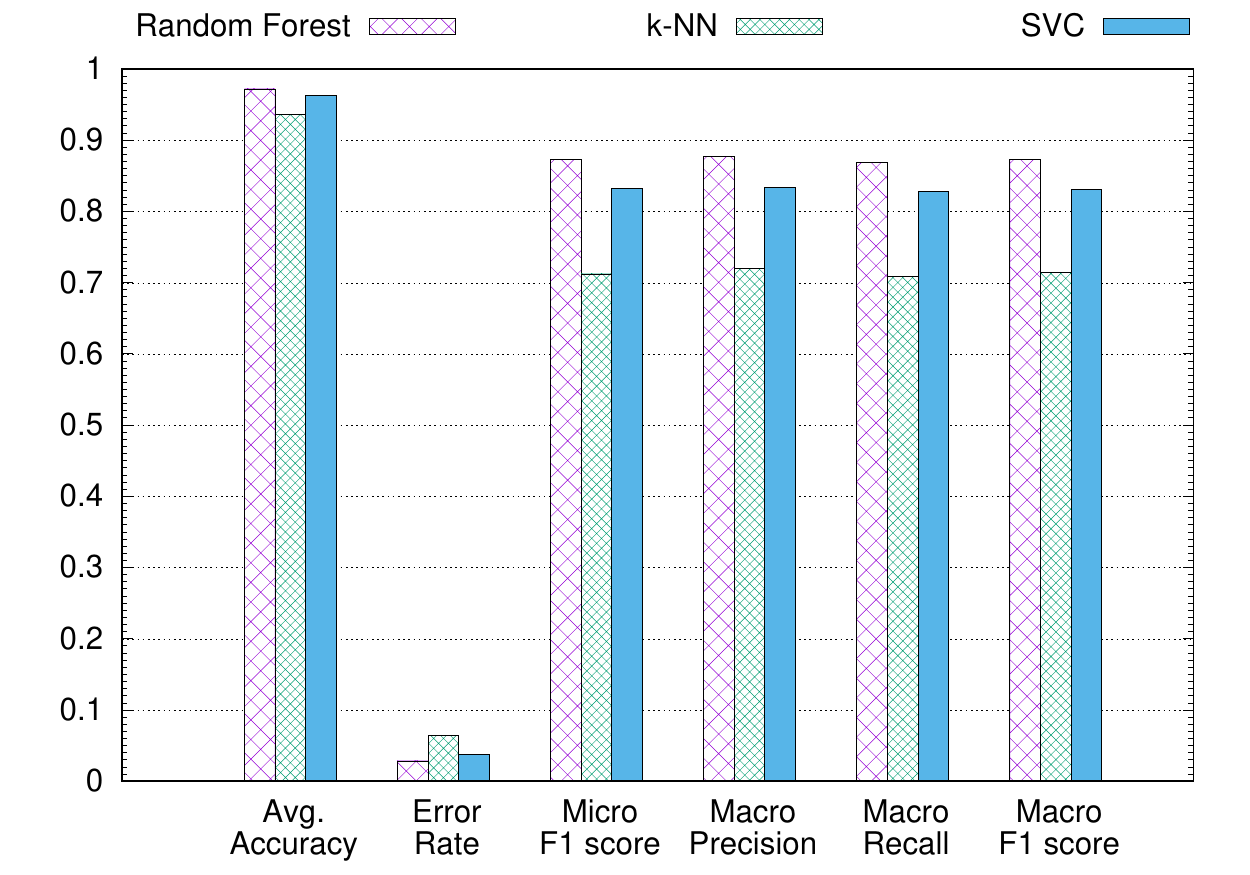}
      \caption{Classifiers' overall performance for Experiment~14.}\label{fig:overall-res-exp14}
      \Description{Classifiers' overall performance for Experiment 14.}
  \end{minipage}
  
  \begin{minipage}[c]{0.5\textwidth}
    \includegraphics[width=\linewidth]{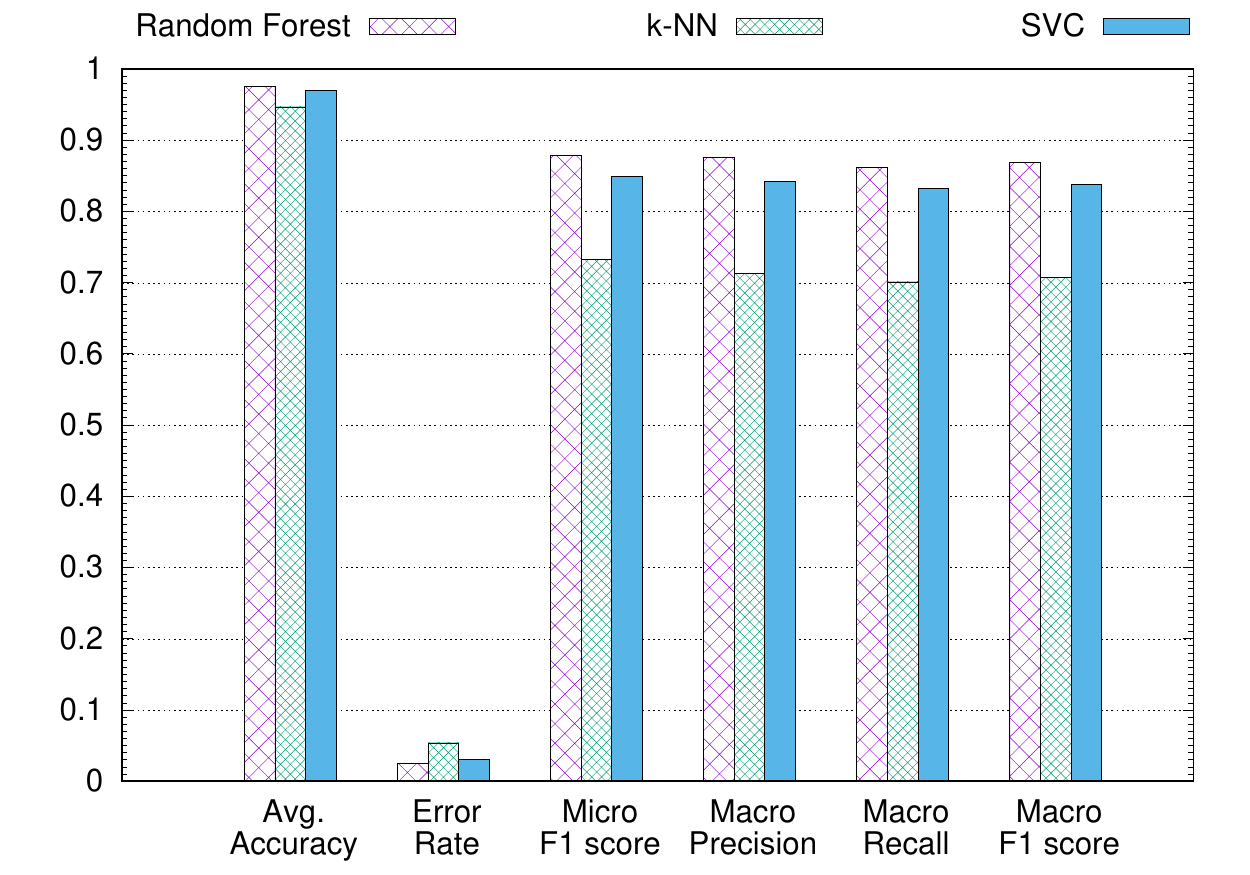}
      \caption{Classifiers' overall performance for Experiment~15.}\label{fig:overall-res-exp15}
      \Description{Classifiers' overall performance for Experiment 15.}
  \end{minipage}%
  \begin{minipage}[c]{0.5\textwidth}
    \includegraphics[width=\linewidth]{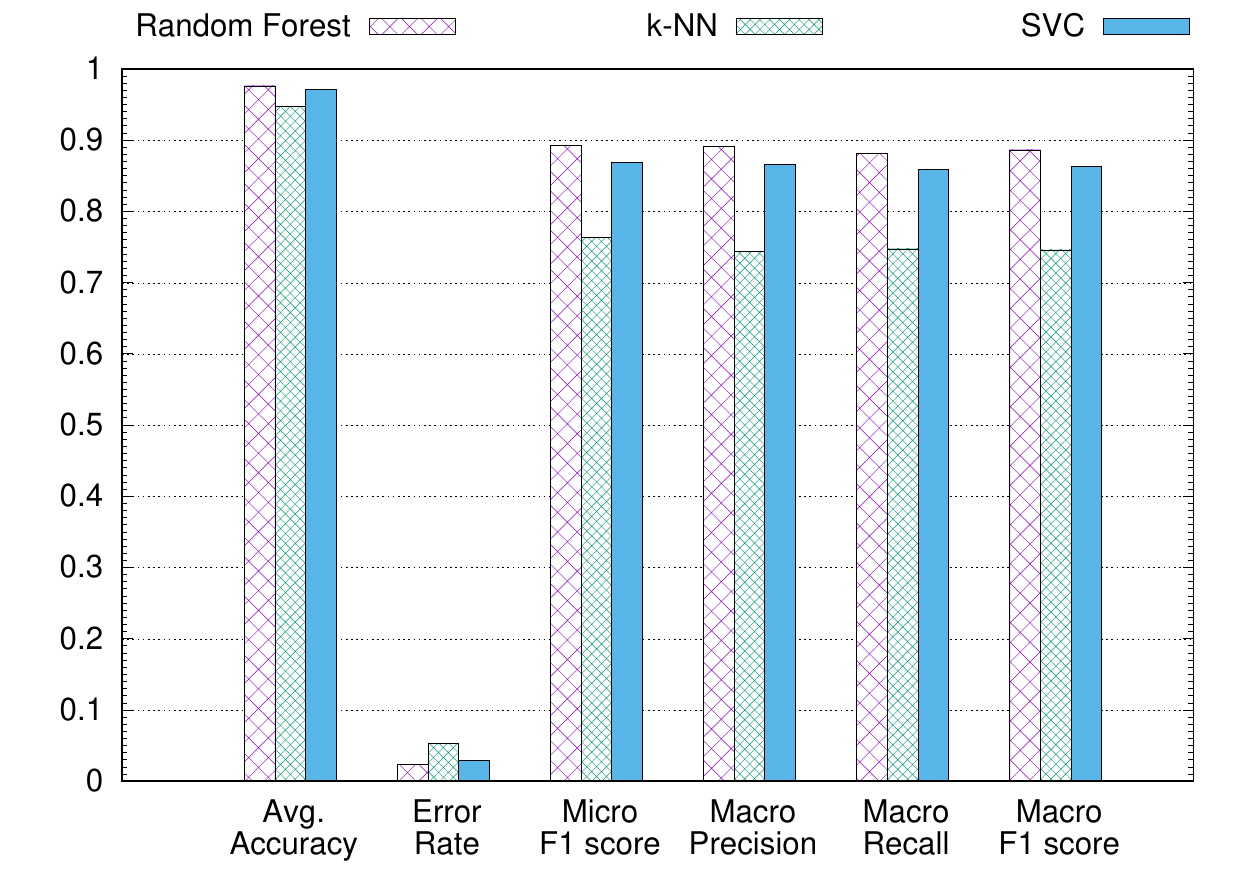}
      \caption{Classifiers' overall performance for Experiment~16.}\label{fig:overall-res-exp16}
      \Description{Classifiers' overall performance for Experiment 16.}
  \end{minipage}
\end{figure*}

This section reports the results of Experiments~5-16 which complements the results of the Experiments~1-4 discussed in section~\ref{sec:results}.
The results are summarized by Figures~\ref{fig:overall-res-exp5}-\ref{fig:overall-res-exp16}.

